\documentclass[onecolumn,floatfix,superscriptaddress,a4paper,showpacs,showkeys,nofootinbib,preprint]{revtex4-1}
\usepackage[colorlinks=true,linktocpage=true,linkcolor=blue,citecolor=blue,allcolors=blue]{hyperref}
\usepackage{epsfig}
\usepackage{latexsym}
\usepackage[utf8]{inputenc}
\usepackage{xspace}
\usepackage{indentfirst}
\usepackage{enumerate}
\usepackage{color}

\usepackage{amsmath}
\usepackage{amssymb}
\usepackage[english]{babel}
\usepackage{url}

\newcommand{\rom}[1]{\uppercase\expandafter{\romannumeral #1\relax}}

\newcommand\tab[2]{
\begin{center}
\begin{table}
\begin{tabular}#1
\end{tabular}
\caption{#2}
\end{table}
\end{center}}

\newcommand{\eq}[1]{\begin{align} #1 \end{align}}

\newcommand{\skewn}[0]{S\sigma}
\newcommand{\kurt}[0]{\kappa\sigma^2}

\begin{document}

\title{Chemical freeze-out conditions and
fluctuations of conserved charges in heavy-ion collisions within quantum van der Waals model}

\author{R. V. Poberezhnyuk}
\affiliation{Bogolyubov Institute for Theoretical Physics, 03680 Kiev, Ukraine}
\affiliation{Frankfurt Institute for Advanced Studies, Giersch Science Center,
D-60438 Frankfurt am Main, Germany}
\author{V. Vovchenko}
\affiliation{Institut f\"ur Theoretische Physik,
Goethe Universit\"at Frankfurt, D-60438 Frankfurt am Main, Germany}
\affiliation{Frankfurt Institute for Advanced Studies, Giersch Science Center,
D-60438 Frankfurt am Main, Germany}
\author{A.~Motornenko}
\affiliation{Institut f\"ur Theoretische Physik,
Goethe Universit\"at Frankfurt, D-60438 Frankfurt am Main, Germany}
\affiliation{Frankfurt Institute for Advanced Studies, Giersch Science Center,
D-60438 Frankfurt am Main, Germany}
\author{M. I. Gorenstein}
\affiliation{Bogolyubov Institute for Theoretical Physics, 03680 Kiev, Ukraine}
\affiliation{Frankfurt Institute for Advanced Studies, Giersch Science Center, D-60438 Frankfurt am Main, Germany}

\author{H. Stoecker}
\affiliation{Institut f\"ur Theoretische Physik,
Goethe Universit\"at Frankfurt, D-60438 Frankfurt am Main, Germany}
\affiliation{Frankfurt Institute for Advanced Studies, Giersch Science Center,
D-60438 Frankfurt am Main, Germany}
\affiliation{GSI Helmholtzzentrum f\"ur Schwerionenforschung GmbH, D-64291 Darmstadt, Germany}

\date{\today}

\begin{abstract}
The chemical freeze-out parameters in central nucleus-nucleus collisions are extracted consistently from hadron yield data within the quantum van der Waals (QvdW) hadron resonance gas model.
The beam energy dependences for skewness and kurtosis
of net baryon,  net electric, and net strangeness charges
are predicted.
The QvdW interactions in asymmetric matter, $Q/B \neq 0.5$, between (anti)baryons yield a non-congruent liquid-gas phase transition, together with a nuclear critical point (CP) with critical temperature of $T_c=19.5$ MeV.
The nuclear CP yields
the collision energy dependence of the skewness and the kurtosis to both
deviate significantly from the ideal hadron resonance gas baseline predictions even far away, in $(T,\mu_B)$-plane, from the CP. These predictions can
readily be tested by  STAR and NA61/SHINE Collaborations at the RHIC BNL
and the SPS CERN, respectively, and by HADES at GSI. The results presented here offer a broad opportunity for the search for signals of phase transition in dense hadronic matter at the future NICA and FAIR high intensity facilities. 
\end{abstract}


\maketitle

\section{Introduction}
The structure of the phase diagram of strongly interacting matter is
one of the most important and still open topics in nuclear and particle physics to date.
The known phenomenology of the physics of strong interactions suggests both, short-range repulsion and intermediate-range attraction between nucleons in proximity of nuclear saturation density $n_0=0.16$ fm$^{-3}$.
This yields a first-order liquid-gas phase transition (LGPT) from a dilute (gaseous) to a dense (liquid) phase of nuclear matter, which smoothens out in the nuclear critical point (CP).
In contrast to the hypothetical deconfinement-related CP,  the existence of the LGPT and the nuclear CP is better established \cite{Pochodzalla:1995xy,
Jennings:1982wi,Ropke:1982ino,Fai:1982zk,Biro:1981es,Stoecker:1981za,Csernai:1984hf,Bondorf:1985mv,Bondorf:1985mp,Molitoris:1986pp,Hahn:1986mb,Aichelin:1988me,Peilert:1989kr,Peilert:1991sm,Bondorf:1995ua}, see~\cite{Borderie:2019fii} for a review. 

Theoretical arguments suggest the enhancement of certain
fluctuations of conserved quantities in the critical region
\cite{Stephanov:1998dy,Stephanov:1999zu,Athanasiou:2010kw,Stephanov:2008qz,Kitazawa:2012at,Vovchenko:2015uda},  namely, the fluctuation of the conserved charges that are related to the so-called order parameter.
The signals of the CP in the scaled variance of the charge fluctuations fade-out rather quickly when moving away from the CP \cite{Vovchenko:2015vxa,Poberezhnyuk:2018mwt}. On the other hand, the CP signals in fluctuation measures which are related to the higher order moments of charge distributions, namely skewness and kurtosis of charge fluctuations, can be seen even far away from the location of the CP on the phase diagram \cite{Vovchenko:2015pya,Vovchenko:2017ayq,Vovchenko:2016rkn}.
Thus, the observed large
deviations of the higher order charge fluctuations from the ideal hadron resonance gas (IHRG) baseline can be taken as a signal for the existence of a CP.

Here we study this issue by employing the quantum van der Waals hadron resonance gas (QvdW-HRG) model, which is the extension of the classical vdW model: The QvdW model was recently generalized to include the grand canonical ensemble (GCE) \cite{Vovchenko:2015xja}, the effects of relativity and quantum statistics \cite{Vovchenko:2015pya}, and the full known spectrum of hadrons and resonances \cite{Vovchenko:2016rkn}. 
The QvdW-HRG model is a minimal interaction-extension of the IHRG model. It takes into account both, attractive and repulsive, interactions between only baryons and between only anti-baryons. These interactions yield the LGPT and the nuclear CP within the model~\cite{Vovchenko:2015pya}. The model includes two parameters only, which are fixed by the properties of the nuclear ground state.

The QvdW-HRG at low temperatures is reduced to normal nuclear matter, described by the QvdW model, see Refs.~\cite{Vovchenko:2015vxa,Vovchenko:2015xja,Vovchenko:2015pya,Redlich:2016dpb,
Vovchenko:2017cbu,Vovchenko:2017ayq}.
The results for symmetric nuclear matter are similar to the Walecka model results~\cite{Poberezhnyuk:2017yhx}. The QvdW model was applied to describe asymmetric nuclear matter and its non-congruent LGPT in Ref.~\cite{Poberezhnyuk:2018mwt}.

The skewness and the kurtosis of baryonic charge fluctuations were calculated within the QvdW-HRG model for central nucleus-nucleus (A+A) collisions along the chemical freeze-out line in Ref.~\cite{Vovchenko:2017ayq}. The present paper extends these results in two directions. First, the chemical freeze-out line is derived consistently for central A+A collisions within the QvdW-HRG model. Second, both the baryonic and electric charge fluctuations are calculated in $T-\mu_B$ plane and along the freeze-out line. The electric charge is a more convenient quantity for experimental measurement, compared to baryonic charge, as it does not require the detection of the dominant electrically neutral baryons. 
The Thermal-FIST \cite{Vovchenko:2019pjl} package is used for the calculations within the QvdW-HRG model.

The paper is organized as follows.
Section \ref{sec-model} briefly describes the QvdW-HRG model.
Section~\ref{sec-pt} discusses the non-congruent LGPT in asymmetric nuclear matter within the QvdW-HRG model and presents chemical freeze-out lines obtained within the QvdW-HRG and IHRG models.
Section~\ref{sec-fluct} presents the QvdW-HRG and IHRG results on the skewness and the kurtosis of charge fluctuations as functions of the collision energy and in the coordinates of baryochemical potential and temperature.
A summary  closes the article in Sec.~\ref{sec-sum}.

\section{The Quantum van der Waals - hadron resonance gas model}
\label{sec-model}

The total baryon ($B$), electric ($Q$), and strangeness ($S$) charges of the hot, dense, hadronic system in the GCE are regulated by the corresponding chemical potentials, $\mu_B$, $\mu_Q$, and $\mu_S$. The chemical potential of the $j$-th type hadron is $\mu_j=b_j\mu_B+s_j\mu_S+q_j\mu_Q$, where $b_j$, $s_j$, and $q_j$ are, respectively, the baryonic number, the strangeness, and the electric charge of the  hadron of $j$ type.
The QvdW model yields the total pressure of the system as a sum of the partial pressures of baryons, anti-baryons, and mesons~\cite{Vovchenko:2016rkn}:
\eq{\label{qvdw-p}
p(T,\mu) = p_B + p_{\bar B} + p_M~. 
}
The partial pressure of the baryons is given as
\begin{eqnarray}
 p_B\left(T,\mu \right) = \sum_{j\in B}p_{j}^{id}\left( T,\mu _{j}^{B \ast }\right)-an_{B}^{2}~. \label{pB}
\end{eqnarray}%
Here $T$ is the temperature, $p_{j}^{\rm id}$ is the ideal Fermi-Dirac pressure of the baryons of $j$ type,  $\mu _{j}^{B \ast }$ and $n_B$ are, respectively, the shifted baryonic chemical potential of  baryons of $j$ type and the total density of all baryons:
\eq{\label{muB*}
\mu _j^{B*}
& = \mu_j - b\,p_{B} - a\,b\,n_{B}^2 + 2\,a\,n_{B}~,\\
n_{B} & = \left[\frac{\partial p_B}{\partial \mu_B}\right]_{T}=\sum_{j \in B}n_j =
\left( 1-bn_B\right) \sum_{j\in
B} n_{j}^{id}\left( T,\mu_j^{B*}\right)~.\label{nB}
}
The corresponding expressions for $p_{\bar B}$, $\mu_j^{\bar{B}*}$, and $n_{\bar{B}}$ of the antibaryons are analogous
to Eqs.~(\ref{pB})-(\ref{nB}). 
The QvdW interactions are assumed to exist separately between all pairs of baryons, 
and between all pairs of antibaryons,
where the same parameters are used for all (anti-)baryons 
as for nucleons, $\ a=329$ MeV fm$^{3}$ and $b=3.42$ fm$^{3}$~\cite{Vovchenko:2016rkn}. These parameters $a$ and $b$ 
were obtained in Ref.~\cite{Vovchenko:2015vxa} by fitting the
saturation density, $n^{\rm GS}=0.16$ fm$^{-3}$, and binding energy, $E_b^{\rm GS}=-16$ MeV,
of the ground state of symmetric nuclear matter. 
Possible QvdW interactions for baryon-antibaryon, meson-meson, and
meson-(anti-)baryon pairs are neglected.
The partial pressure of all mesons is taken as a sum of the ideal Bose-Einstein gas pressures.
The summation in Eqs.~(\ref{pB})-(\ref{nB}) is performed over all hadrons and resonances listed in the Particle Data Tables \cite{Patrignani:2016xqp} and which have a confirmed status there.
\begin{figure}[ht!]
\includegraphics[width=0.55\textwidth]{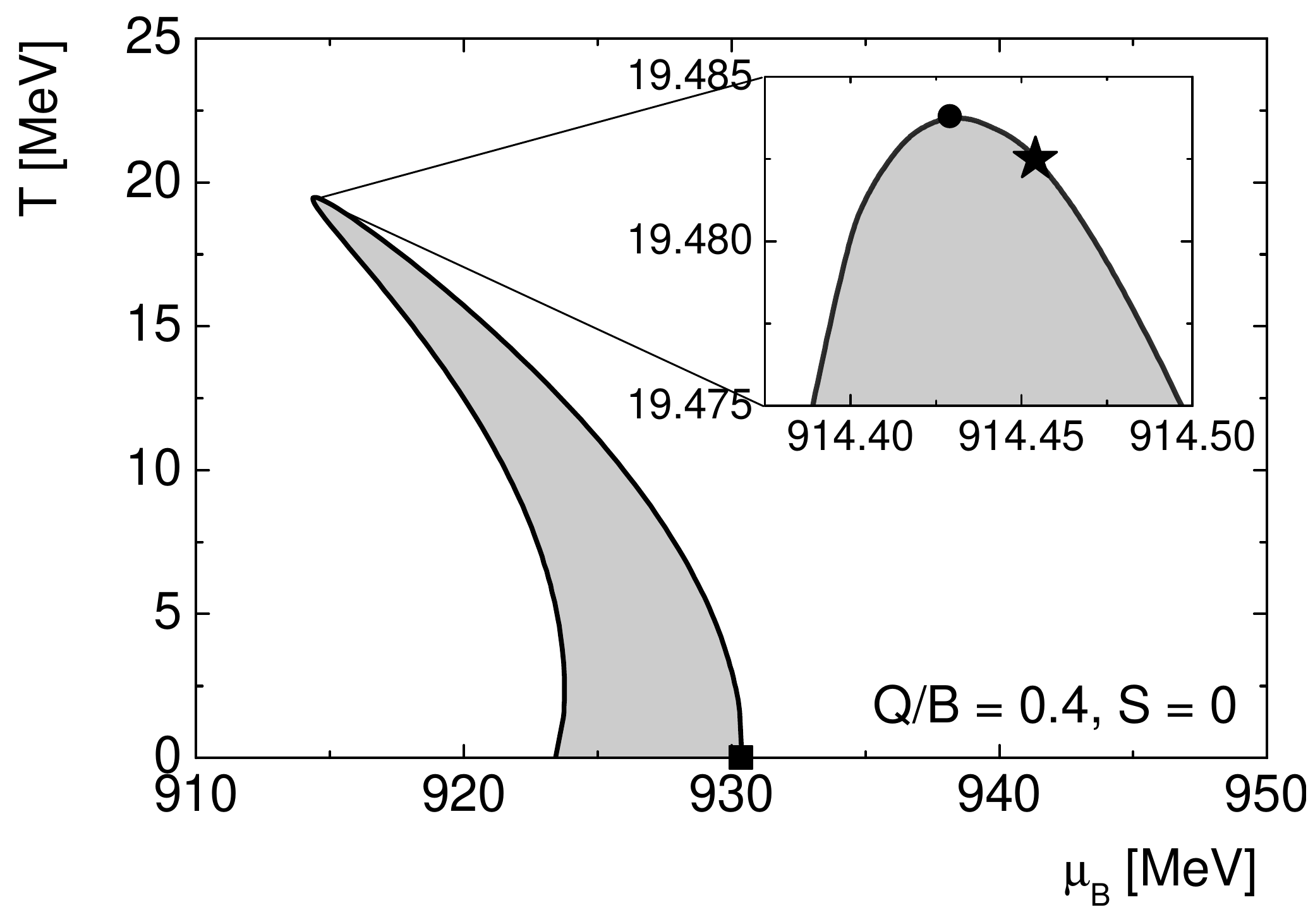}
\caption{\label{pt} 
Liquid-gas phase transition for asymmetric nuclear matter with an asymmetry parameter of $Q/B=0.4$ in the ($\mu_B,~T$) coordinates. The shaded area represents the mixed phase. The ground state is denoted by a square. The inset shows a zoomed-in picture of the region around the critical point. The critical point and the temperature endpoint are shown by the star and the circle, respectively. 
}
\end{figure}

\section{phase transition and chemical freeze-out}
\label{sec-pt}

The LGPT in the QvdW-HRG model is due to an interplay of the repulsive and the attractive interactions. The mixed phase boundary and the location of the CP in asymmetric nuclear matter are found from the Gibbs equilibrium condition \cite{Poberezhnyuk:2018mwt}.
The QvdW-HRG model with $Q/B=0.4$ exhibits the CP at $\mu_B^c=914.5 ~{\rm MeV}$, $T_c=19.48 ~{\rm MeV}$. 
The LGPT region in the ($\mu_B$,~T) coordinates is shown in Fig. \ref{pt}.

We fix the ratio of the electric-to-baryon charge (asymmetry parameter): $Q/B=0.4$. This approximately corresponds to the isospin asymmetry in heavy nuclei like lead~(Pb) or gold~(Au).
Note that isospin asymmetry alters qualitatively the properties of the PT, rendering it as being a ``non-congruent'' PT.
As a result, the 
mixed phase in ($\mu_B$,~T) coordinates can not be presented by a line, but is rather a region of finite width. Moreover, for a non-congruent PT, the location of the CP differs from the location of the temperature endpoint~(TEP), the point with the maximum temperature at which the phase coexistence is possible. The inset in Fig.~\ref{pt} zooms in on the region of the CP. The star and the circle represent the CP and the TEP, respectively,  see Refs.~\cite{Poberezhnyuk:2018mwt,Yang:2019lwx} for details.

The particle number fluctuations in A+A collisions are calculated
within statistical models
at different collision energies by using the chemical freeze-out values of the temperature and
baryochemical potential.
In the present paper we use the data on mean hadron multiplicities in various experiments at SchwerIonen-Synchrotron
(SIS), Alternating Gradient Synchrotron (AGS), Super
Proton Synchrotron (SPS), and Large Hadron Collider (LHC)
to determine the chemical freeze-out values for the QvdW-HRG model. 
\tab{
{| l | c | c | c | c | c | c | c | c | c |}
\hline
\hline
 $\sqrt{s_{NN}}$ & Refs. & \multicolumn{4}{|c}{IHRG} & \multicolumn{4}{|c|}{QvdW-HRG} \\
\cline{3-10}
  ${\rm [GeV]}$  &  & $\mu_B$ [MeV] & T [MeV] & $V$ $[{\rm fm}^{3}]$ & $\gamma_s$ & $\mu_B$ [MeV] & T [MeV] & $V$ $[{\rm fm}^{3}]$& $\gamma_s$ \\
 \hline
 $2760$ & \cite{Abelev:2013vea,Abelev:2013xaa,ABELEV:2013zaa,Abelev:2014uua}& 1.4$\pm$9.7 & 153$\pm$3 &4741$\pm$540& 1.11$\pm$0.03 & 1.5$\pm$10.9 & 156$\pm$5 &4148$\pm$613& 1.10$\pm$0.03\\
  $17.3$ & \cite{Afanasiev:2002mx,Alt:2006dk,Alt:2008qm,Alt:2008iv,Alt:2004kq,Anticic:2011ny,Friese:2002re,Anticic:2011zr} & 246$\pm$10 & 150$\pm$3  &4885$\pm$522 & 0.87$\pm$0.03& 314$\pm$39 & 160$\pm$7 &3503$\pm$543& 0.81$\pm$0.04\\
  $12.3$ &  \cite{Afanasiev:2002mx,Alt:2006dk,Alt:2008qm,Alt:2008iv,Alt:2004kq}& 289$\pm$14 & 151$\pm$5 &3552$\pm$447& 0.72$\pm$0.04 & 356$\pm$48 & 153$\pm$9 & 3474$\pm$474& 0.69$\pm$0.04\\
  $8.8$ &  \cite{Afanasiev:2002mx,Alt:2006dk,Alt:2008qm,Alt:2008iv,Alt:2004kq}& 372$\pm$12 & 145$\pm$4  &2657$\pm$304& 0.80$\pm$0.04& 443$\pm$46 & 143$\pm$7&3210$\pm$345 & 0.78$\pm$0.04\\
    $7.7$ &  \cite{Alt:2006dk,Alt:2007aa,Alt:2008qm,Alt:2008iv}& 415$\pm$11 & 143$\pm$4  &2210$\pm$268& 0.84$\pm$0.05& 491$\pm$57 & 138$\pm$8 &3093$\pm$369& 0.80$\pm$0.05\\
     $6.3$ &  \cite{Alt:2006dk,Alt:2007aa,Alt:2008qm,Alt:2008iv}& 469$\pm$12 & 138$\pm$7  &1935$\pm$404& 0.82$\pm$0.05& 566$\pm$106 & 131$\pm$10 & 2829$\pm$654& 0.83$\pm$0.05\\
      $4.9$ & \cite{Ahle:1999jm,Ahle:1999in,Ahle:1999va,Becattini:2000jw} & 569$\pm$16 & 120$\pm$4  &2905$\pm$695& 0.70$\pm$0.08& 634$\pm$81 & 119$\pm$8 &2896$\pm$815& 0.70$\pm$0.08\\
         $2.3$ & \cite{Cleymans:1998yb,Averbeck:2000sn} & 808$\pm$25 & 48.3$\pm$2 &---& --- & 802$\pm$23 & 48.3$\pm$2 &---& ---\\
  \hline
}{\label{tab-data}
The results of the hadron chemical freeze-out parameters fits for the IHRG and the QvdW model.
}

The GCE can be used for 
all data sets considered,
except for the lowest energies at SIS. 
The exact net strangeness conservation is enforced for the SIS data, i.e., the calculations are done for these low-energy Au+Au collisions within the strangeness canonical ensemble (SCE)~\cite{BraunMunzinger:2001as,Cleymans:2016qnc}.

Finite resonance widths are treated in the present paper in the framework of energy independent Breit-Wigner scheme~\cite{Vovchenko:2018fmh}. 
Note, that the energy dependent Breit-Wigner scheme leads to a
better description of hadron yields at the LHC \cite{Vovchenko:2018fmh}.
Another possibility is to neglect finite widths of resonances altogether.
Here we stick to the energy independent Breit-Wigner scheme so as to preserve consistency with our earlier works regarding the chemical freeze-out conditions in the IHRG model~\cite{Vovchenko:2015idt} or thermodynamic properties of the QvdW-HRG model~\cite{Vovchenko:2016rkn}.
We did verify that differences in the extracted freeze-out parameters obtained within these different schemes are small, with a possible exception of the strangeness saturation factor $\gamma_S$. 
A detailed study of finite resonance widths effects on hadron yields for intermediate collision energies will be presented elsewhere.

\begin{figure}[h!]
\includegraphics[width=0.49\textwidth]{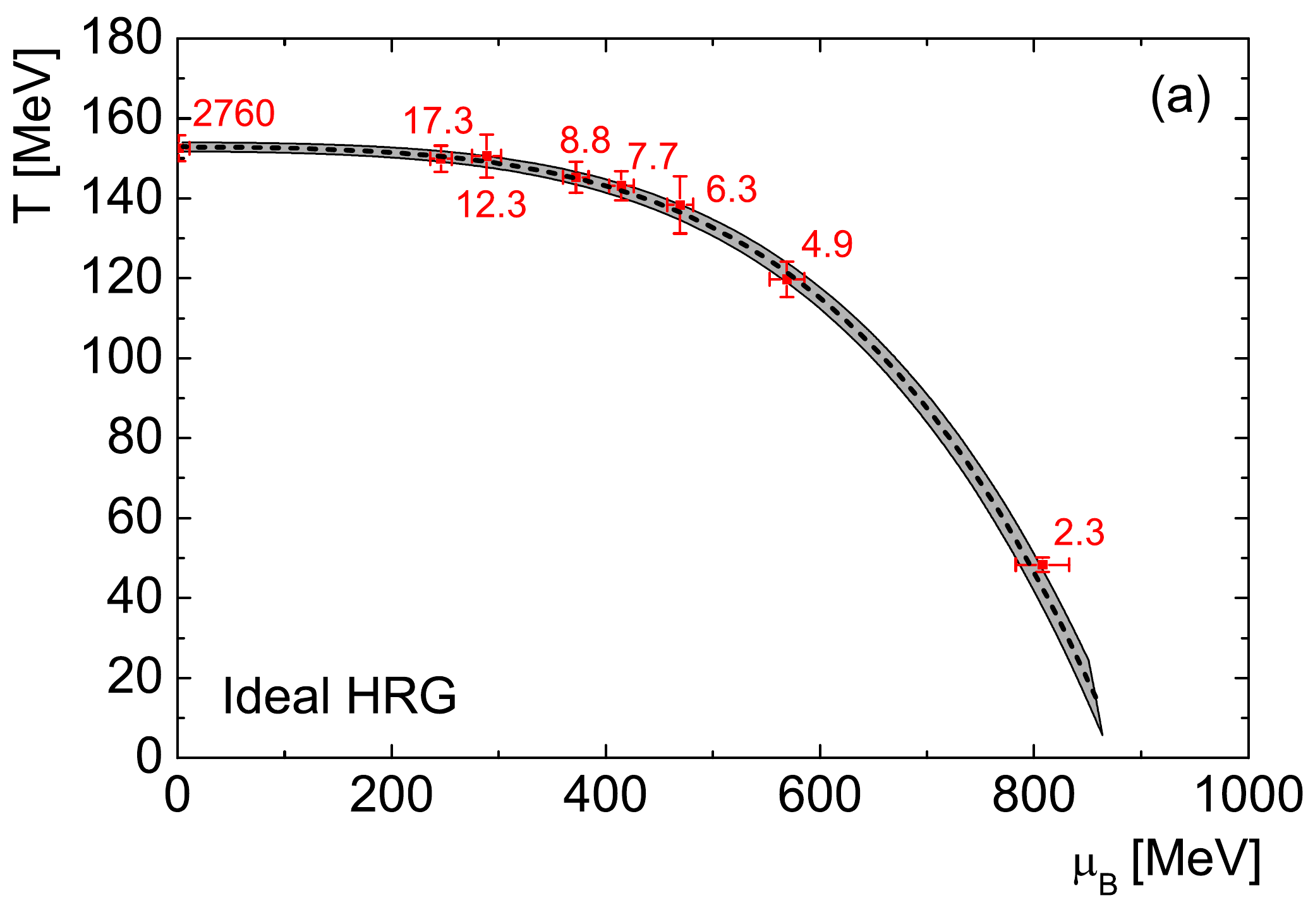}
\includegraphics[width=0.49\textwidth]{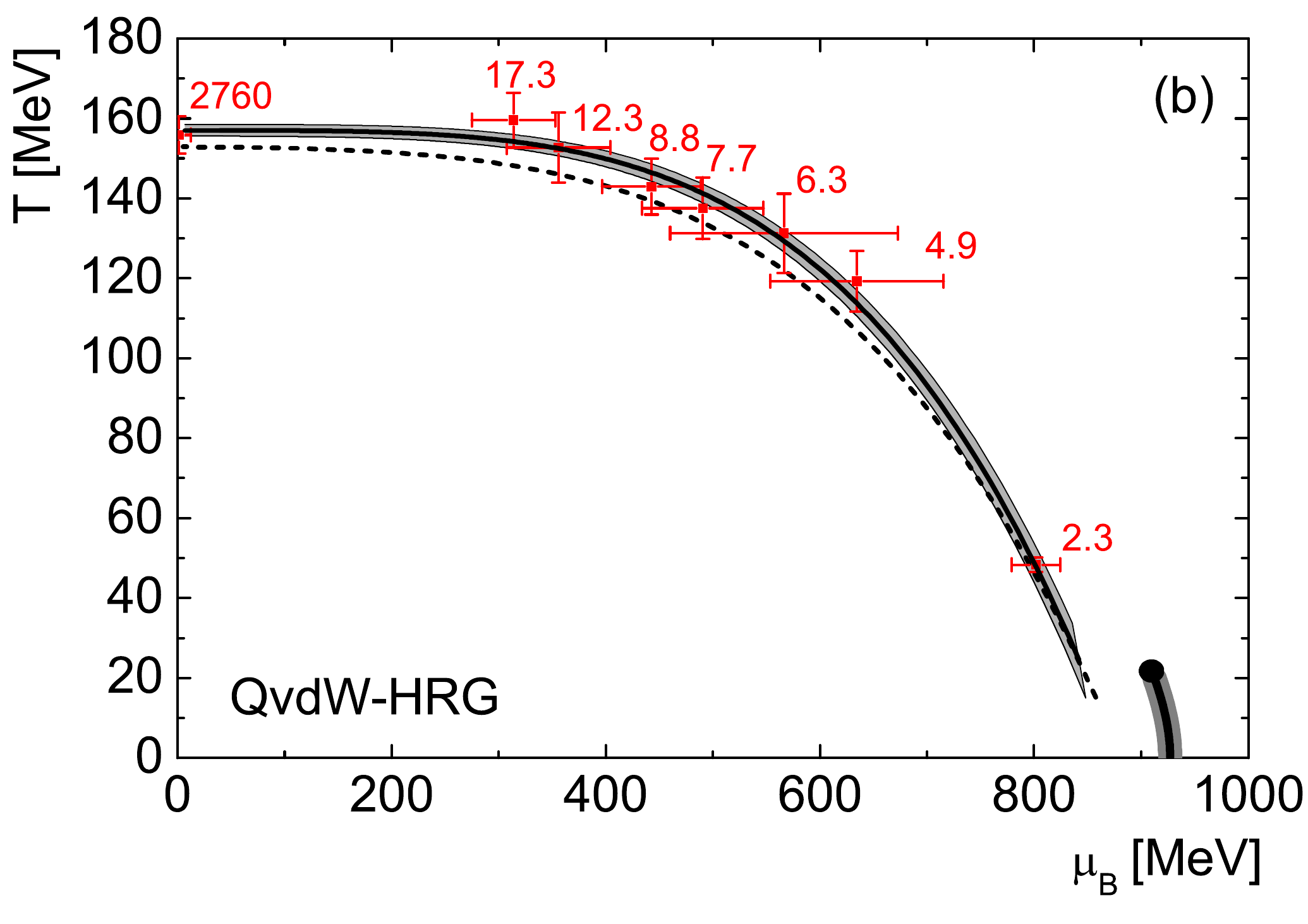}
\caption{\label{fo} 
Freeze-out lines according to Eq.~(\ref{Tmu}) are shown in the ($\mu_B,~ T$) coordinates plane all the way from collision energies $\sqrt{s_{NN}}=1.9~{\rm GeV}$ to $2.8~{\rm TeV}$  for the Ideal-HRG ($a$) and QvdW-HRG model ($b$).  The shaded areas along the curves represent the uncertainties.
The Ideal-HRG freeze-out line is represented in panel ($b$) by dashed line for comparison with the QvdW-HRG curve. The numbers on the freeze-out line give the respective center of mass energy, $\sqrt{s_{NN}}$, in GeV. The first-order phase transition region and the nuclear critical point are also shown in panel (b) by the dark curve and the dot, respectively.
}
\end{figure}

\begin{figure}[h!]
\includegraphics[width=0.49\textwidth]{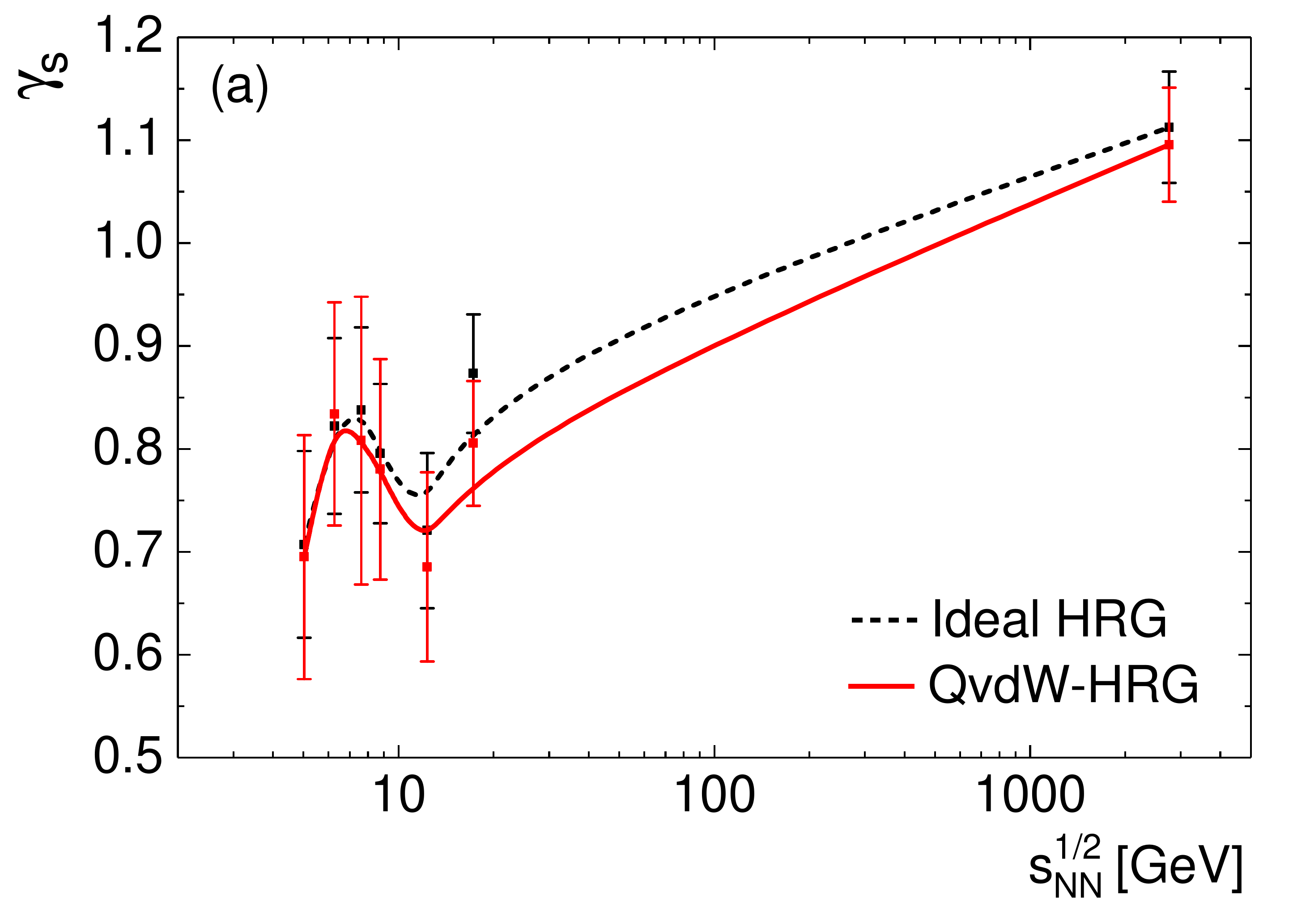}
\includegraphics[width=0.49\textwidth]{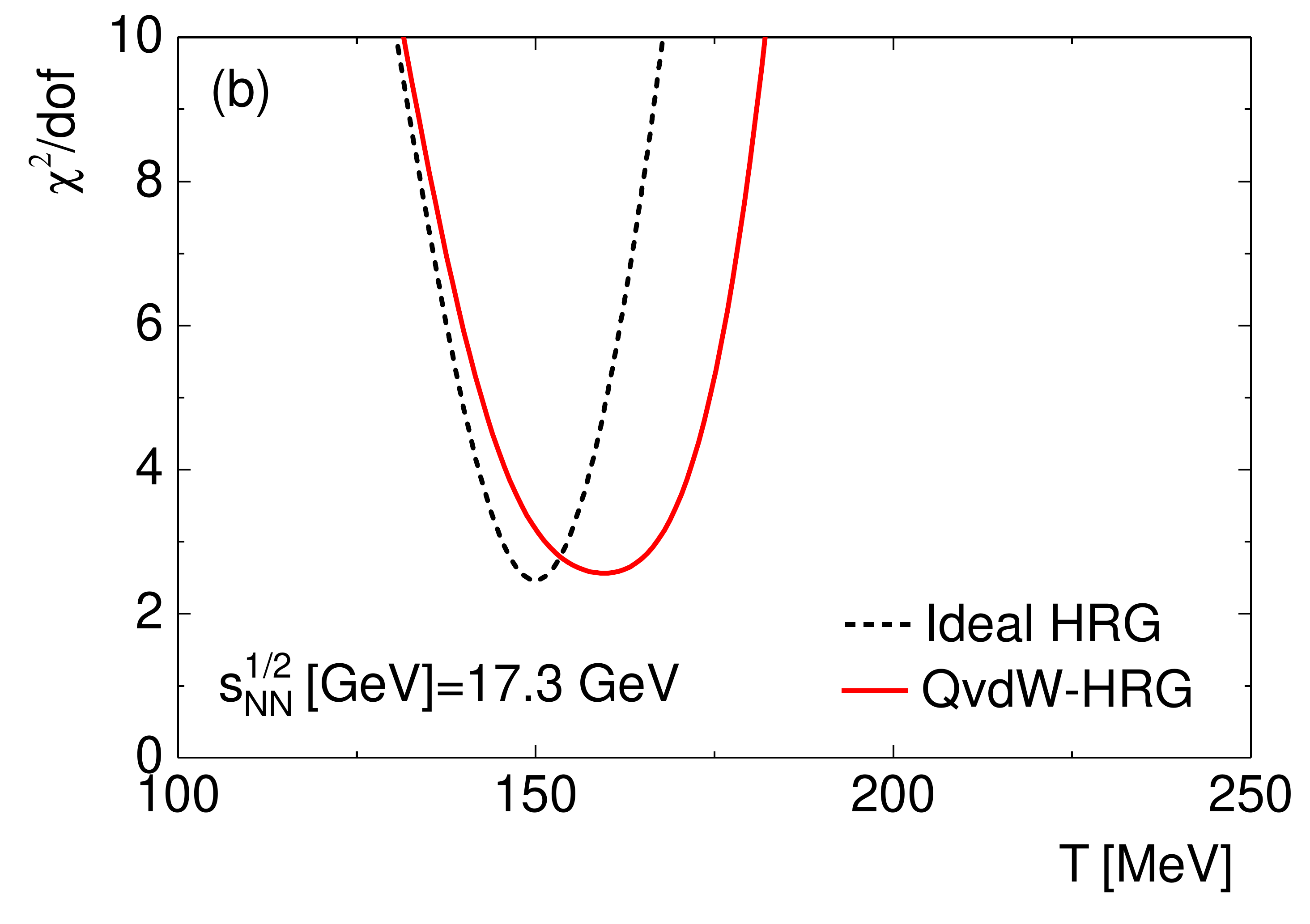}
\caption{\label{gamma} 
(a) $\gamma_S$ is shown along chemical freeze-out line as a function of collision energy.
(b) The temperature dependence of $\chi^2/dof$ of the fits to NA61/SHINE data at  $\sqrt{s_{NN}}=17.3$~GeV using the IHRG (black dashed curves) and QvdW-HRG (red solid curves) model, respectively. Note the higher temperature, $\Delta T \approx 10$~MeV, obtained from the QvdW-HRG fit.
}
\end{figure}

The fitted freeze-out parameters are $\mu_B$, $T$, volume of the system $V$, and the strangeness under-saturation parameter $\gamma_S$ (see Ref.~\cite{Rafelski:2015cxa}).  The corresponding IHRG and QvdW-HRG  fit results for $\mu_B$, $T$, $V$, and $\gamma_s$ are presented in Table \ref{tab-data}.
In the SCE for the SIS data we set the strangeness correlation volume equal to the volume of the system, i.e. $V_c = V$. 
The extracted values of the chemical freeze-out parameters, $T$ and $\mu_B$, are plotted for all energies in Figs.~\ref{fo} (a) and (b) for the IHRG and QvdW-HRG models, respectively. The
values of $\gamma_s$ for both the IHRG and QvdW-HRG models are plotted in Fig.~\ref{gamma} (a). 
The Au+Au data at SIS allow to extract both the temperature and the baryochemical
potential. They are shown in Figs.~\ref{fo} (a) and (b). The
parameter $\gamma_S$ however cannot be reliably determined (see Ref.~\cite{Vovchenko:2015idt}), Hence,
$\gamma_S$ is not shown in Fig.~\ref{gamma} (a) at
SIS.
We define uncertainties of the extracted $\mu_B$, $T$, and $\gamma_s$ values following the procedures given in Ref.~\cite{Becattini:2003wp},
by multiplying the uncertainties inferred from the $\chi^2 = \chi^2_{\rm min} + 1$ contours by a factor $\sqrt{\chi^2_{\rm min}/dof}$~\cite{Patrignani:2016xqp}.

The higher temperatures and larger uncertainties of $T$ and $\mu_B$ values in the QvdW-HRG model are mainly due to the 
excluded volume interactions of (anti-)baryons which is incorporated in this model.
Such an effect has been reported before for different variants of the excluded volume HRG model~\cite{Vovchenko:2015cbk,Vovchenko:2016ebv}.
Moreover, an appearance of peculiar two-minima structures in $\chi^2$ are possible when deviations from the IHRG model picture are considered~\cite{Satarov:2016peb}, although physics interpretation of the 2nd minimum at higher temperatures is challenging.
For the QvdW-HRG model used here we do observe a second minimum in the $\chi^2$ temperature profile of the ALICE data fit at $T \sim 200$~MeV, but we do not observe any two-minima structures for all other data sets~(SPS and SIS) used in our analysis.
Figure~\ref{gamma}~(b) depicts the temperature profile of $\chi^2/dof$ of the fit to the NA49 data at top SPS energy ($\sqrt{s_{NN}}=17.3$ GeV), the picture for all other energies~(except LHC) is similar.
This figure illustrates the broadening of the $\chi^2$ profile when QvdW interactions between baryons are switched on.

We adopt a simple thermodynamic parametrization of the chemical freeze-out line,
\eq{
T =a_{1}-a_{2}\mu_B^{2}-a_{3}\mu_B^{4}~,~~~~
\mu_B=\frac{b_{1}}{1+b_{2}\sqrt{s_{NN}}}~, \label{Tmu}
}
used previously in Ref.~\cite{Cleymans:2005xv}.
Here we use it to parameterize the extracted $T$ and $\mu_B$ values.
The five newly fitted 
parameters in Eq.~(\ref{Tmu}), $a_1,~a_2,~a_3,~b_1,~b_2$, 
are presented in Table~\ref{tab1} for both the IHRG and QvdW-HRG model.
Note that the QvdW-HRG parameters $~a_2,~a_3,~b_1$ differ by about 20\% from the IHRG fits, but that the $b_2$ value of the IHRG fit exceeds the QvdW-HRG value by 70\%.
The parametrization~\eqref{Tmu} extrapolates the freeze-out line from $\mu_B \approx 0$ region at the highest collision energies 
down to the nuclear matter region of the phase diagram at the lowest collision energies.
The chemical freeze-out line close to the region of the nuclear liquid-gas transition was previously considered in Refs.~\cite{Bastian:2016xna,Roepke:2017ohb} in a context of light cluster formation.

\tab{
{| l | c | c | c | c | c |}
\hline
\hline
 &~$a_1$ [GeV]~  &~~$a_2$ [GeV$^{-1}$]~~~&~~$a_3$ [GeV$^{-3}$]~~~&~~$b_1$ [GeV]~~~&~~$b_2$ [GeV$^{-1}$]~~~ \\
\hline
 IHRG              & $0.152 \pm 0.001$  & $0.026 \pm 0.003$ & $0.219 \pm 0.004$ & $1.310 \pm 0.006$ & $0.278 \pm 0.003$ \\
QvdW           & $0.157 \pm 0.002$ & $0.0032 \pm 0.0027$ & $0.259 \pm 0.004$ & $1.094 \pm 0.004$ & $0.157 \pm 0.002$ \\
  \hline
   \hline
}{\label{tab1}
The freeze-out line parameters, see Eq.~(\ref{Tmu}), for the Ideal-HRG and the QvdW-HRG models. 
}

\begin{figure}[h!]
\includegraphics[width=0.49\textwidth]{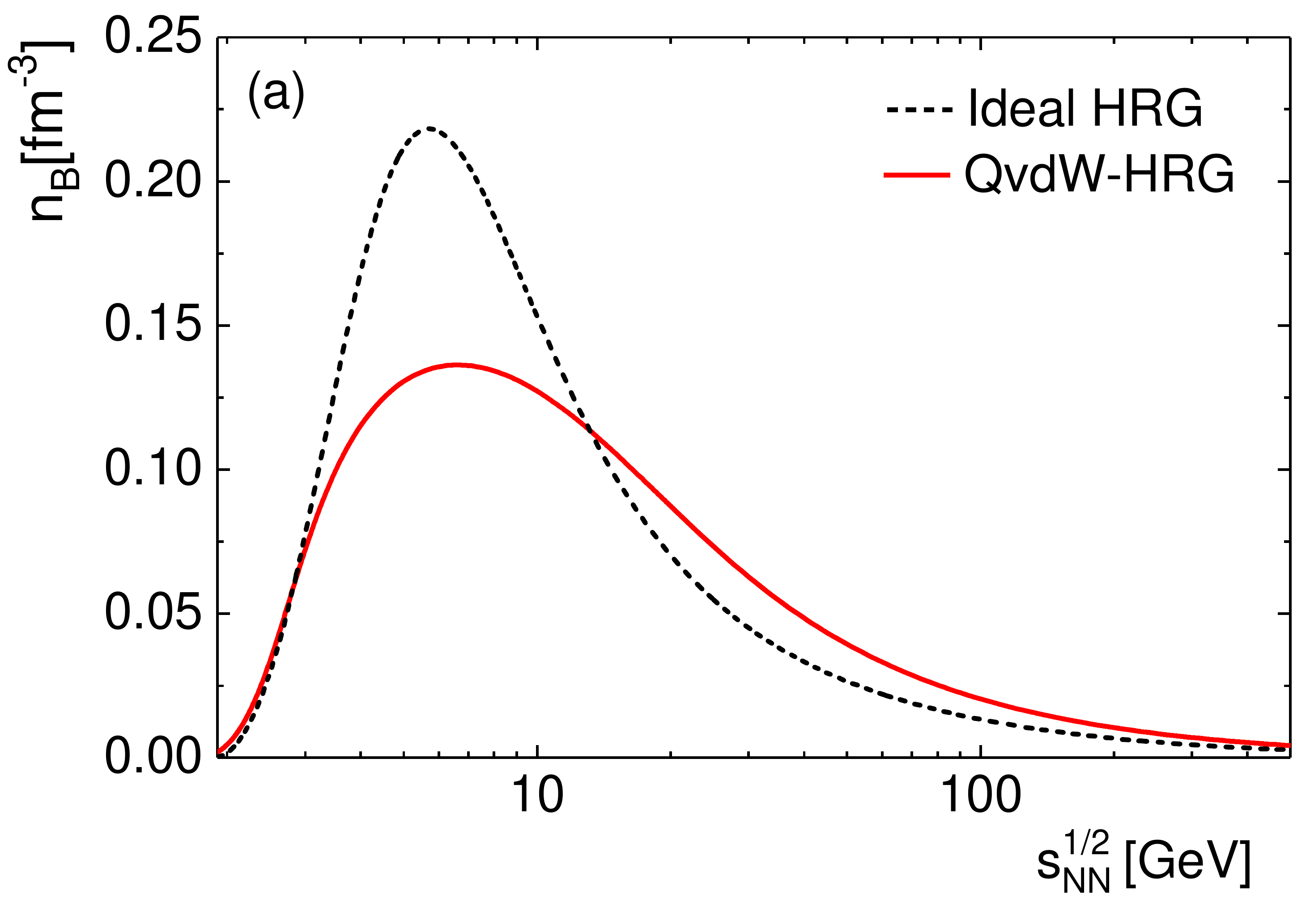}
\includegraphics[width=0.49\textwidth]{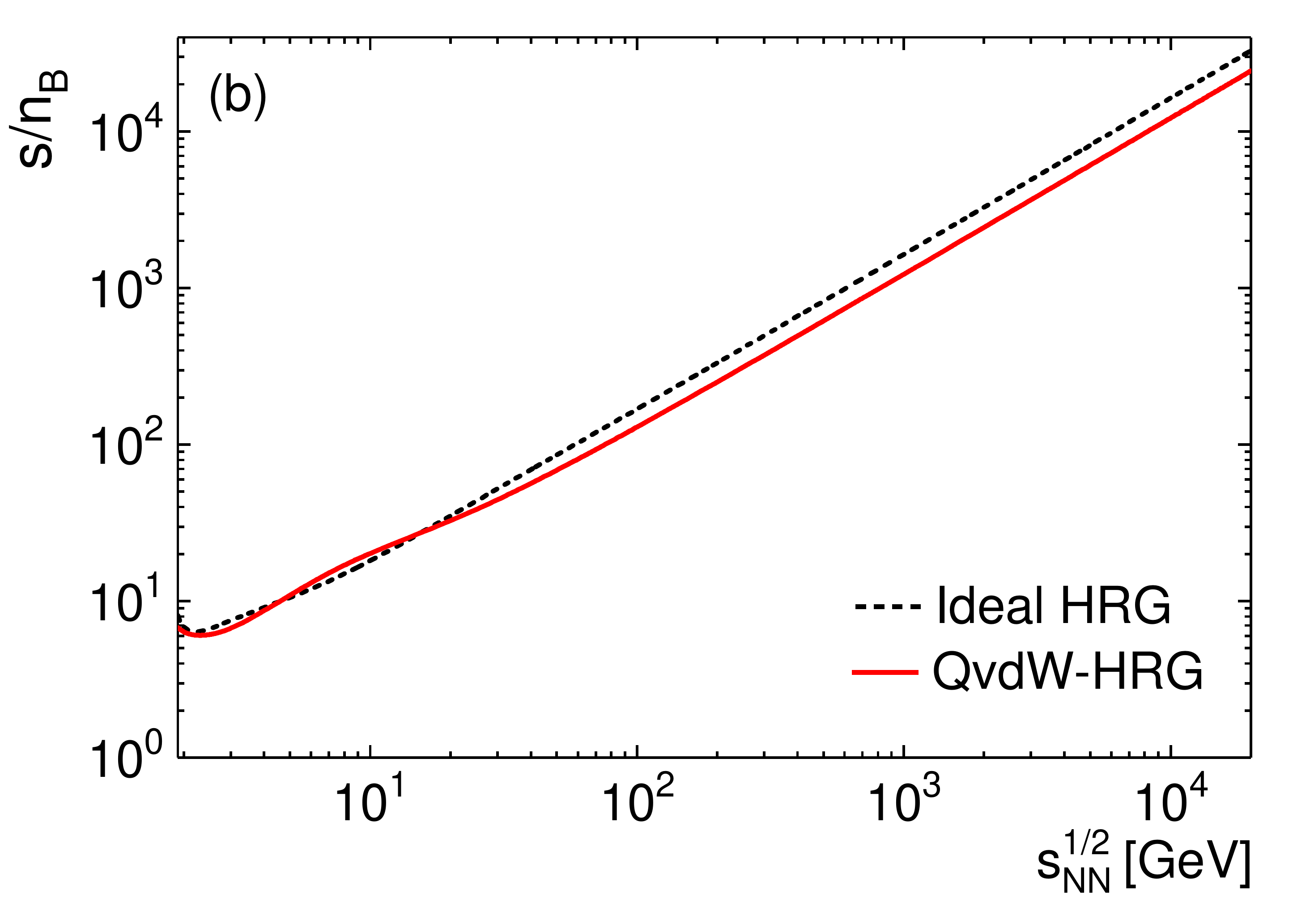}
\caption{\label{fig-entropy1} 
(a) Net baryon density and (b) entropy per baryon are shown along chemical freeze-out line as a function of collision energy. 
}
\end{figure}

Figures \ref{fig-entropy1} (a) and (b) present, respectively, the net baryon density, $n_B$, and the entropy per baryon, $s/n_B$, along the chemical freeze-out line as a function of collision energy for the two models.  At low $\sqrt{s_{NN}}$ in both QvdW-HRG and IHRG models freeze-out takes place in a diluted region.
At the lowest considered energy of $\sqrt{s_{NN}}=1.9$~GeV freeze-out in both models takes place at baryon density of 
$n_B = 2.17 \cdot 10^{-4}$~fm$^{-3}$. 
The net baryon density exhibits a maximum as a function of collision energy, as first noted in Ref.~\cite{Randrup:2006nr} for the IHRG model.
Within the QvdW-HRG model this maximum is lower in comparison to the IHRG model, this is due to the excluded volume repulsion between (anti-)baryons~(see also Ref.~\cite{Begun:2012rf}).
The behavior of the entropy per baryon is very similar for both models, indicating that $s/n_B$ is a robust observable that depends little on the details of the HRG model~\cite{Vovchenko:2016ebv}.

\begin{figure}[h!]
\includegraphics[width=0.49\textwidth]{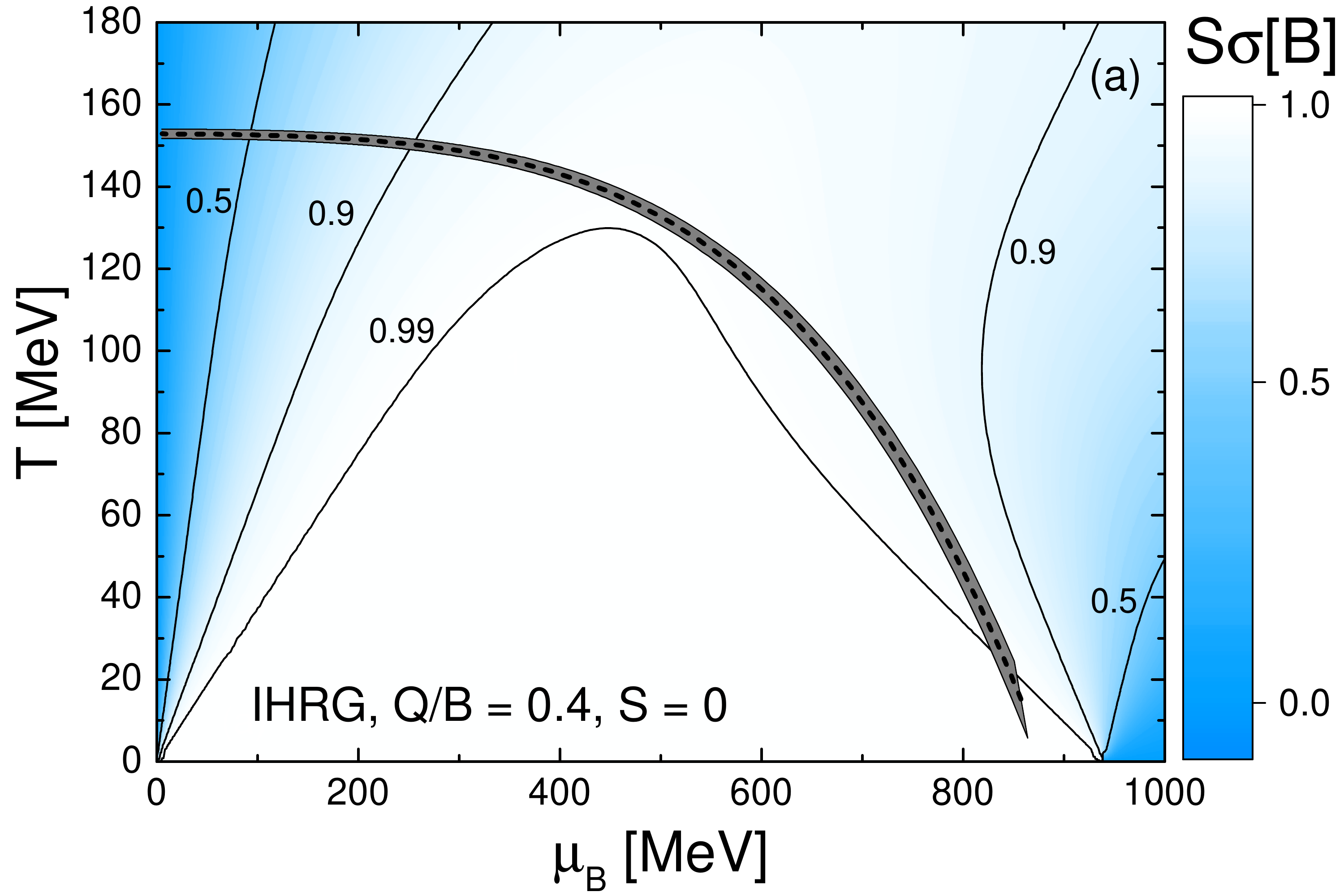}
\includegraphics[width=0.49\textwidth]{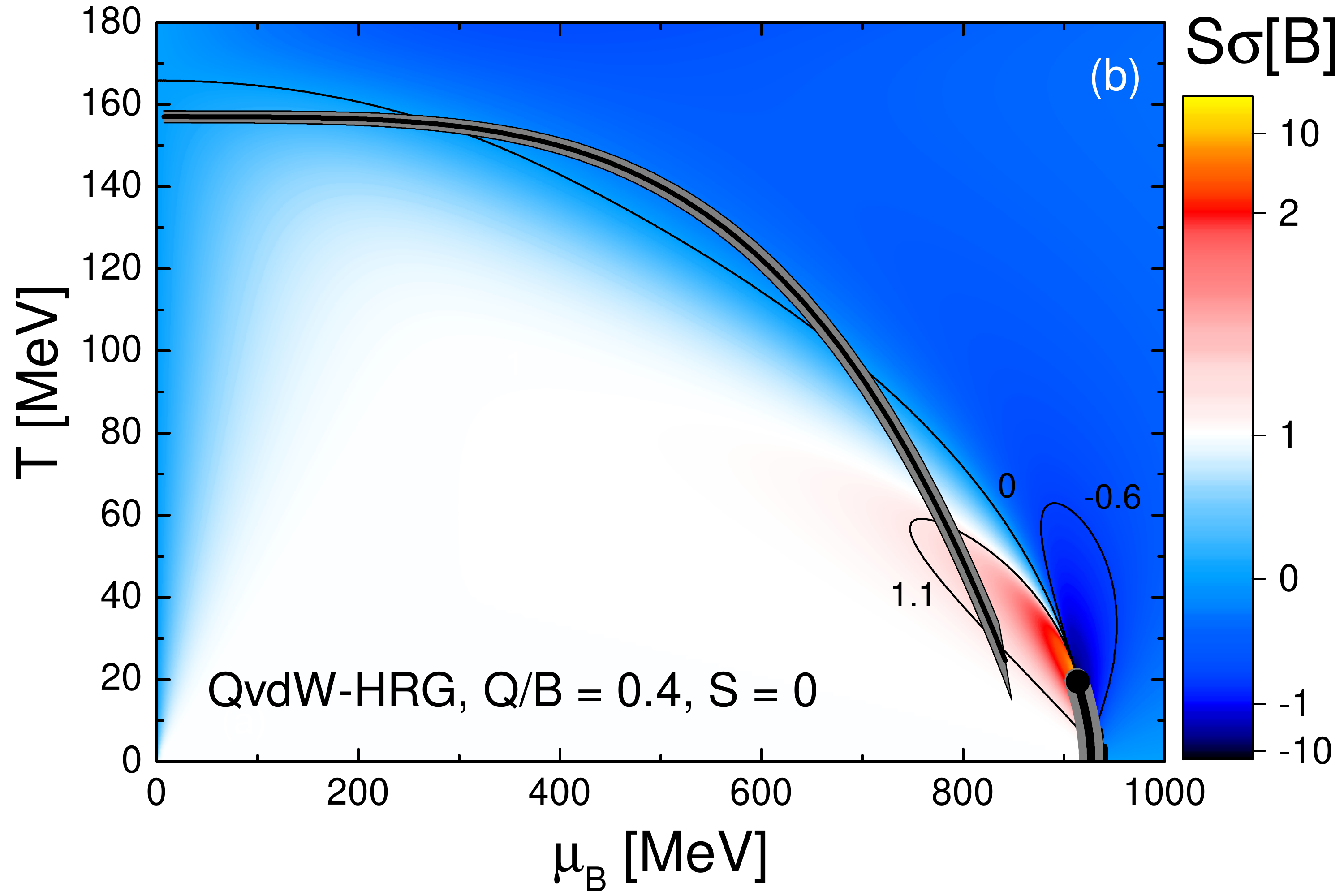}
\includegraphics[width=0.49\textwidth]{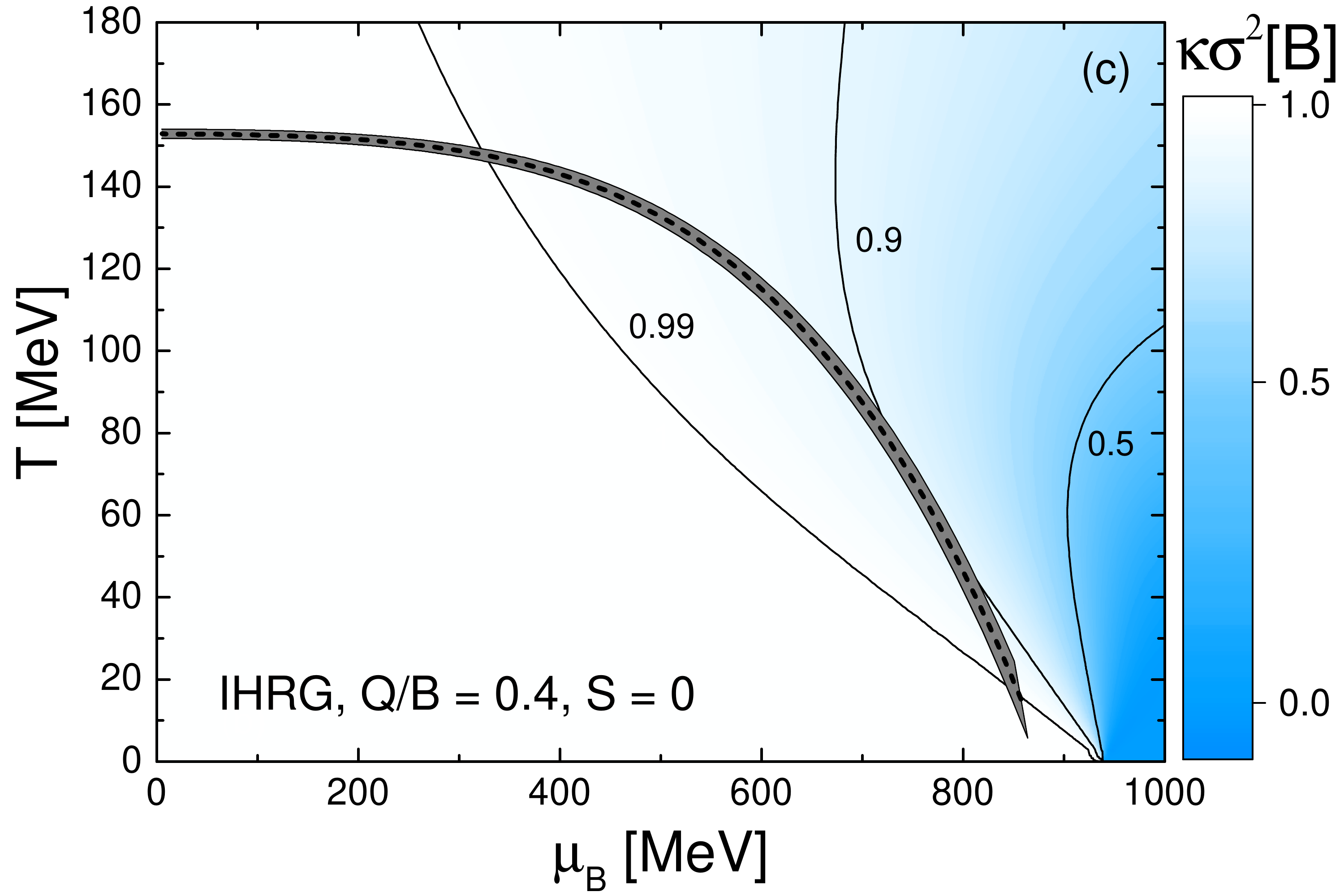}
\includegraphics[width=0.49\textwidth]{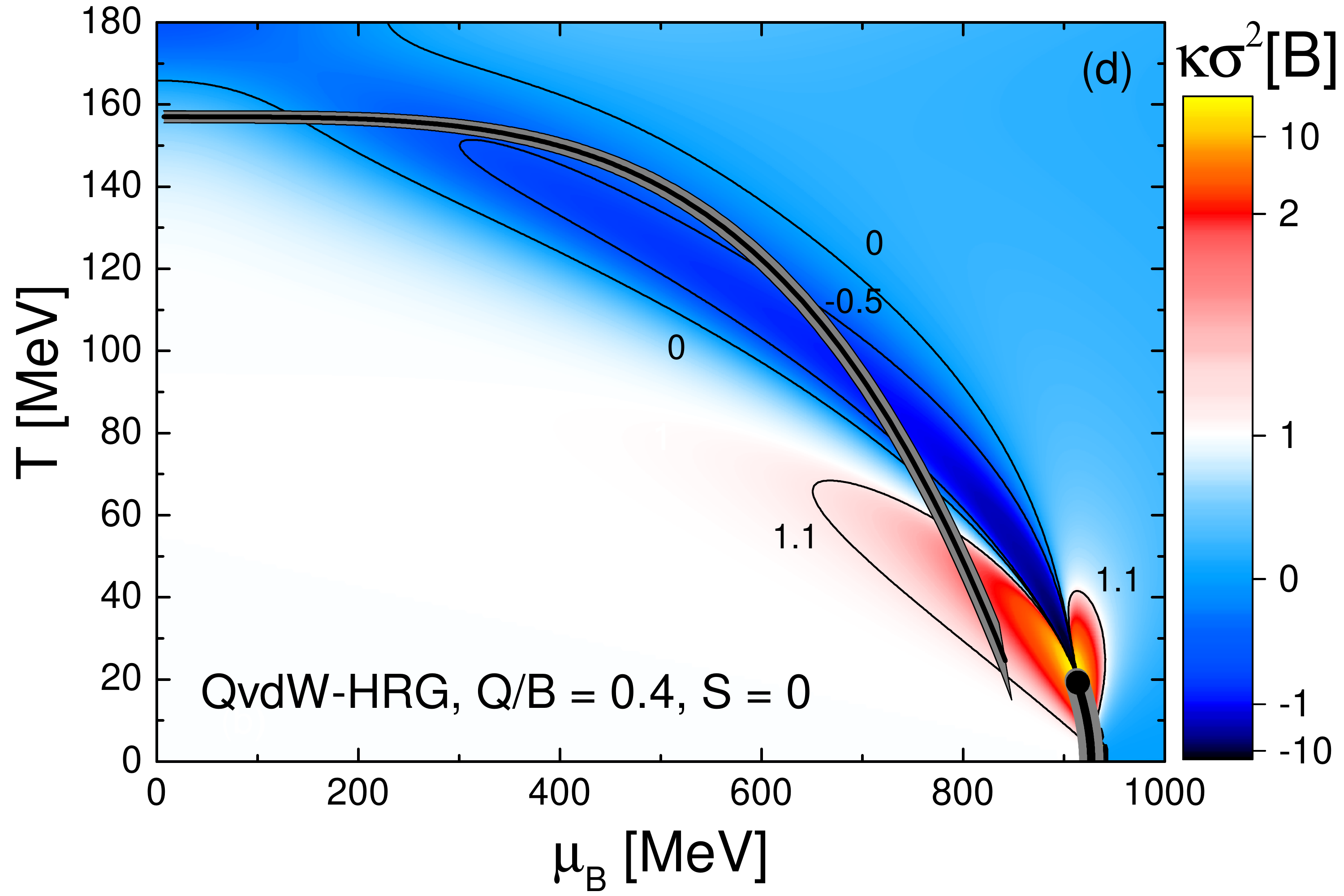}
\caption{\label{baryonic-fluct} 
($a,b$) Skewness, $S\sigma$, and ($c,d$) kurtosis, $\kurt$, of baryonic charge fluctuations in the ($\mu_B,~ T$) coordinates obtained in the IHRG model ($a,c$) and the QvdW-HRG model ($b,d$) for strongly interacting matter with asymmetry parameter $Q/B=0.4$. The freeze-out lines for both models are also shown. Note that there are large differences between the two model predictions for both skewness and kurtosis. 
}
\end{figure}

\begin{figure}[h!]
\includegraphics[width=0.49\textwidth]{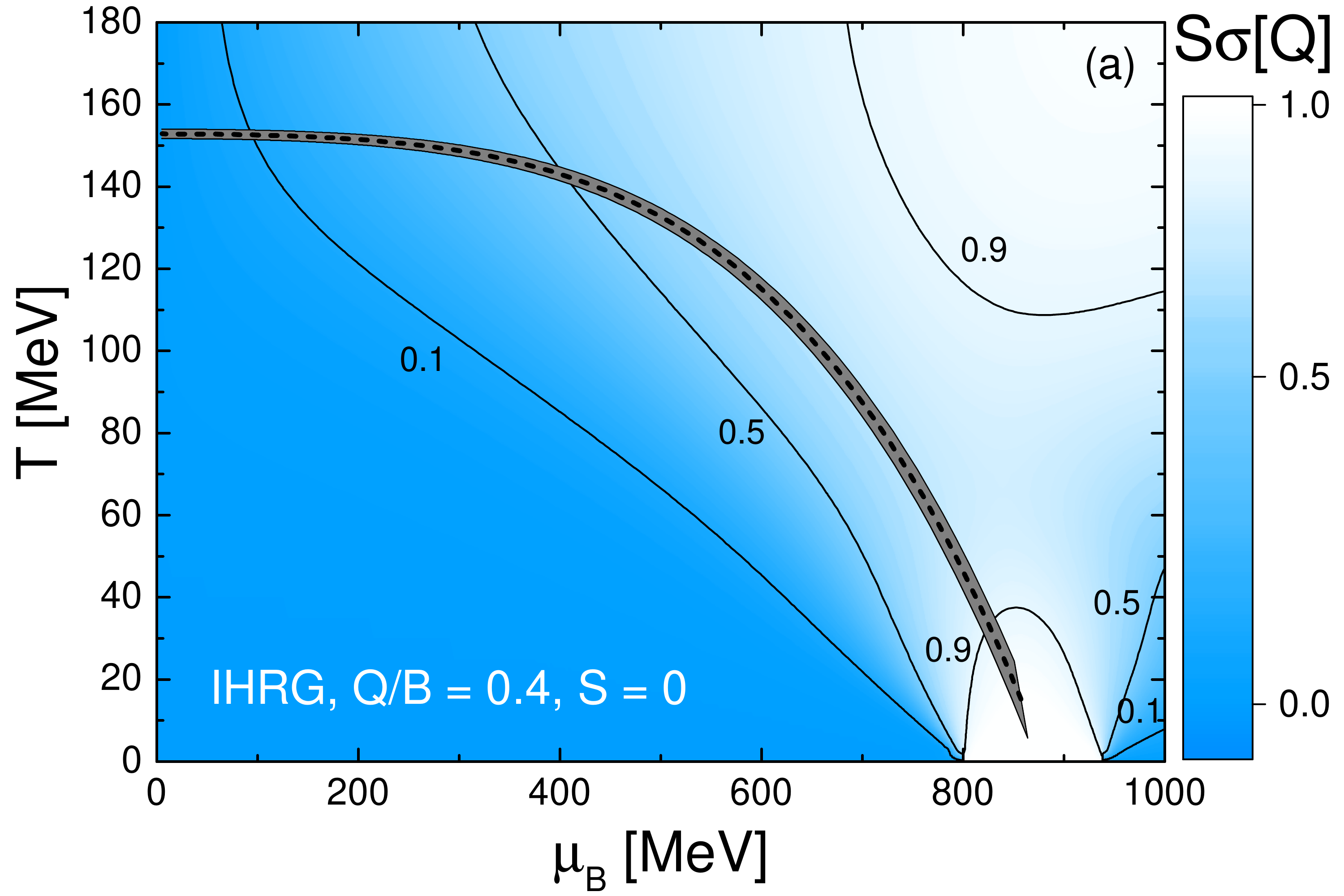}
\includegraphics[width=0.49\textwidth]{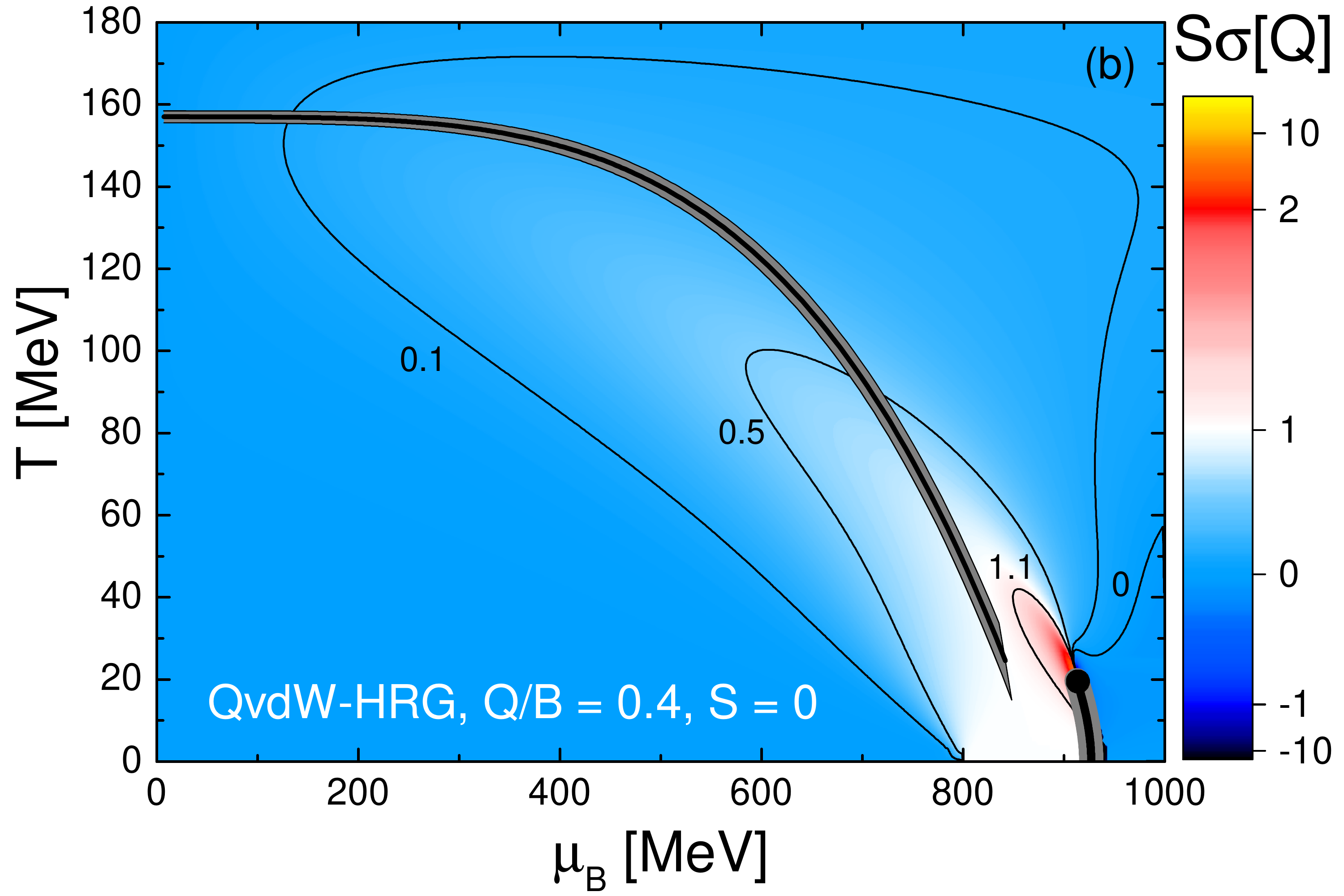}
\includegraphics[width=0.49\textwidth]{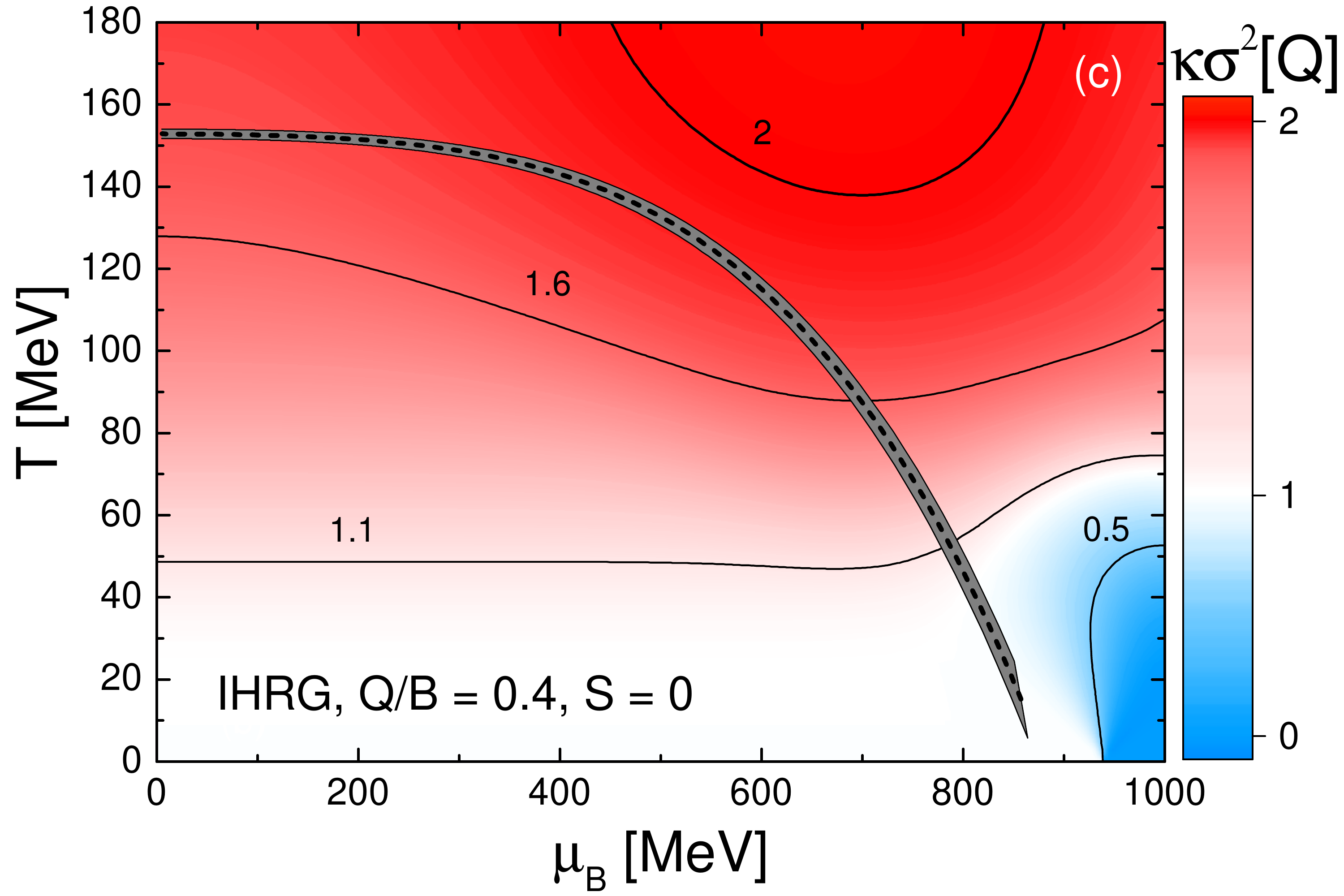}
\includegraphics[width=0.49\textwidth]{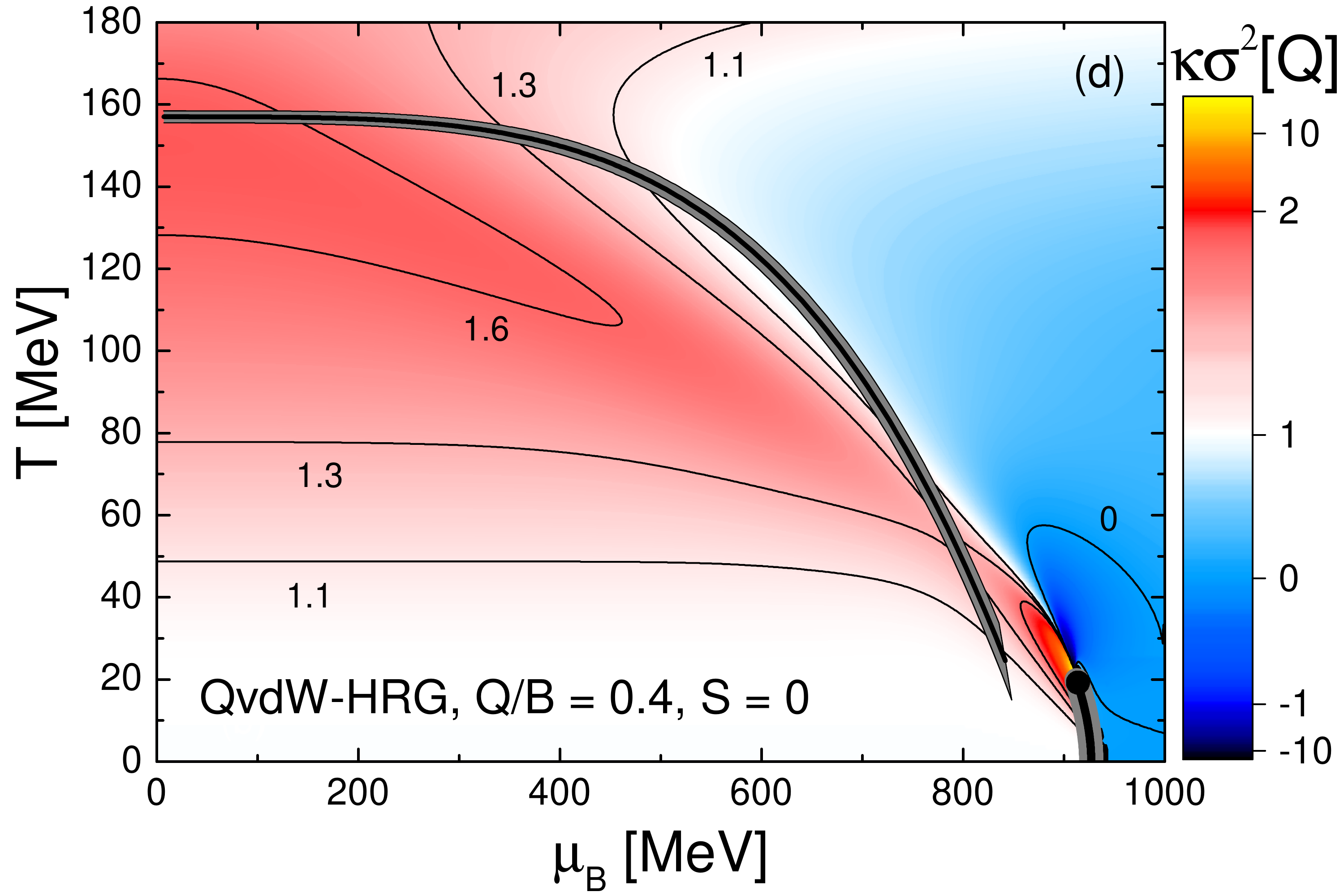}
\caption{\label{electric-fluct} 
The same as in Fig. \ref{baryonic-fluct} but for electric charge fluctuations.
}
\end{figure}

\section{fluctuations}
\label{sec-fluct}

The
skewness, $S\sigma$, and the kurtosis, $\kappa \sigma ^{2}$, of the baryonic, $ch=B$, and the electric, $ch=Q$, charge fluctuations
are expressed as ratios of normalized cumulants (susceptibilities):
\begin{eqnarray}
S\sigma[ch]  =\frac{\chi^{ch}_{3}}{\chi^{ch}_{2}}~,~~~~
\kappa \sigma ^{2}[ch] =\frac{\chi^{ch}_{4}}{\chi^{ch}_{2}}~.
\label{kurt}
\end{eqnarray}
Susceptibilities, $\chi^{ch}_i$,  are calculated in the GCE from the scaled total pressure by taking the derivatives with respect to the corresponding powers of the chemical potentials over the temperature:
\eq{
\chi^{ch}_{n}& =\frac{\partial ^{n}\left( p/T^{4}\right) }{\partial \left( \mu_{ch}
/T\right) ^{n}}~.
}

Figures \ref{baryonic-fluct} and \ref{electric-fluct} show the skewness and the kurtosis of, respectively, the baryonic and the electric charge fluctuations in the ($\mu_B,~ T$) coordinate plane as calculated in the IHRG and the QvdW-HRG models. White coloring corresponds to $\skewn=\kurt=1$.
The third order susceptibility  is anti-symmetric with respect to hadrons and anti-hadrons, $\chi_3^{ch}=\chi_3^{h}-\chi_3^{\bar h}$. At $\mu_B= 0$ the numbers of hadrons and anti-hadrons are equal.
Therefore, at $\mu_B= 0$ the skewness of both charge fluctuations in both models equals zero, $\skewn =0$. 
In contrast, the fourth order susceptibility  is symmetric with respect to hadrons and anti-hadrons, $\chi_4^{ch}=\chi_4^{h}+\chi_4^{\bar h}$. Therefore, $\chi_4$'s of particle and anti-particle fluctuations do not cancel  out each other at $\mu_B= 0$. 
At high temperatures, the contribution of the pions, which are Bose particles, to $\chi_4^Q$ is substantial. 
This brings to the large positive values of $\kurt[Q]\approx 1.6$ at high temperatures. 
\begin{figure}
\includegraphics[width=0.49\textwidth]{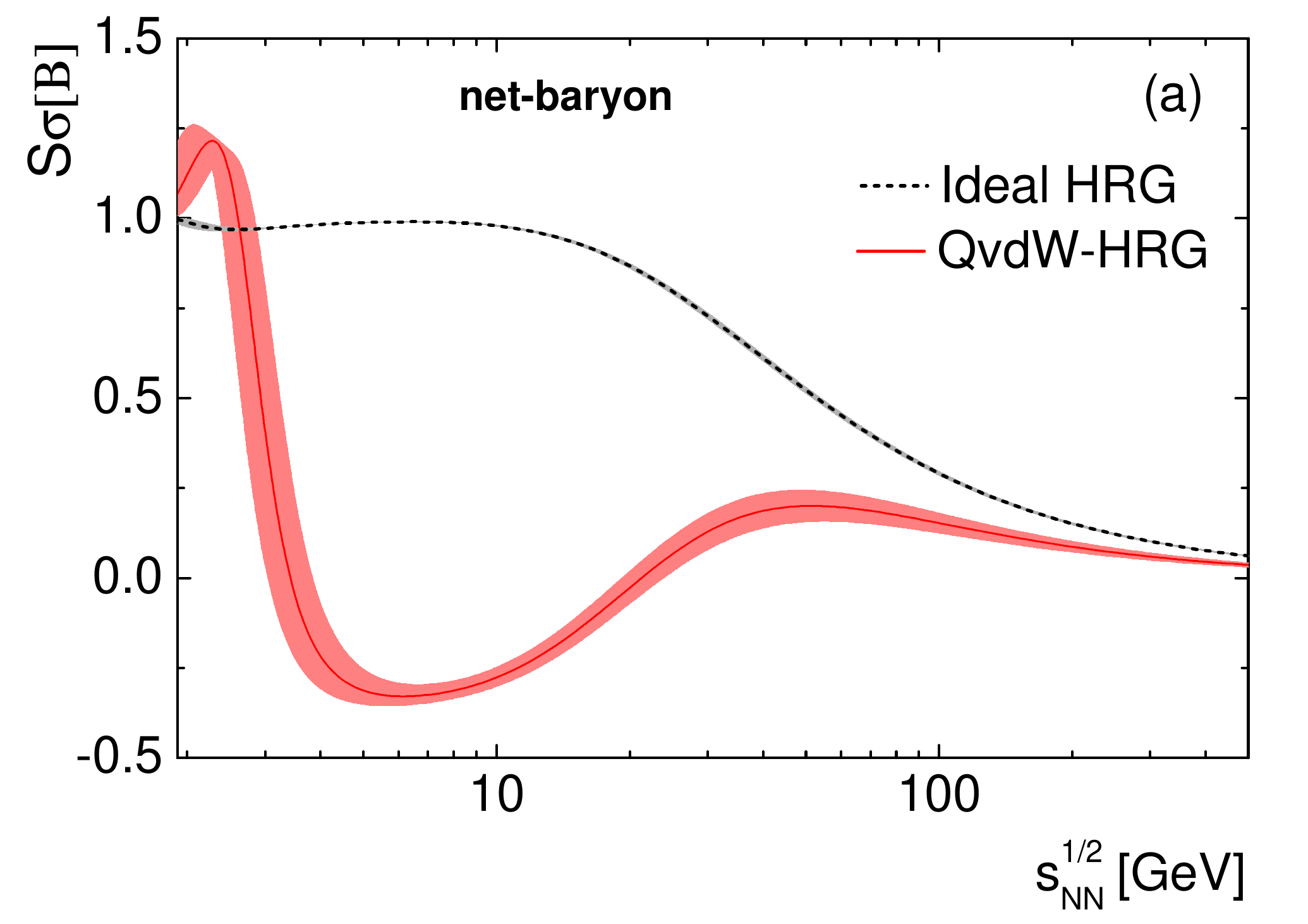}
\includegraphics[width=0.49\textwidth]{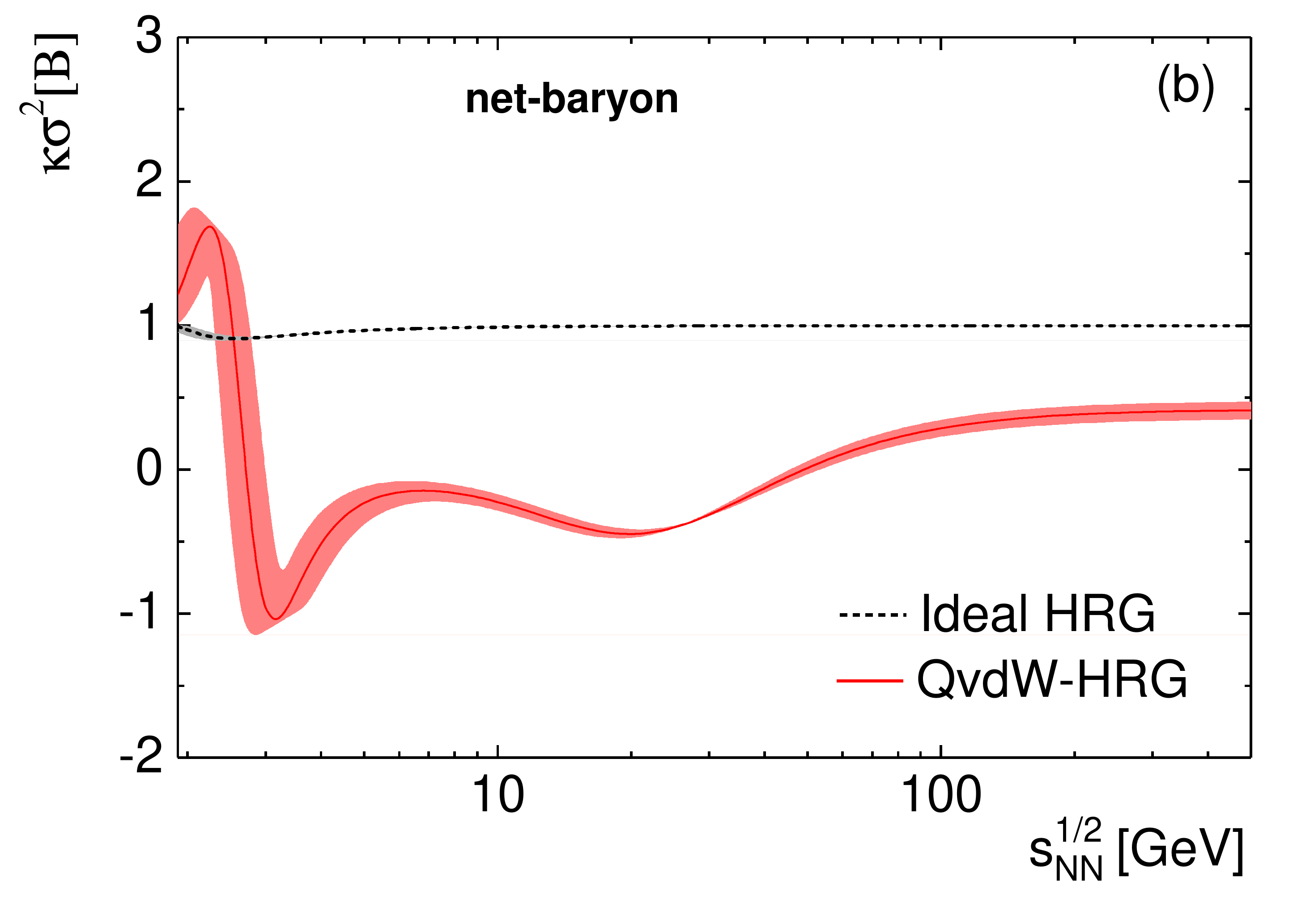}
\includegraphics[width=0.49\textwidth]{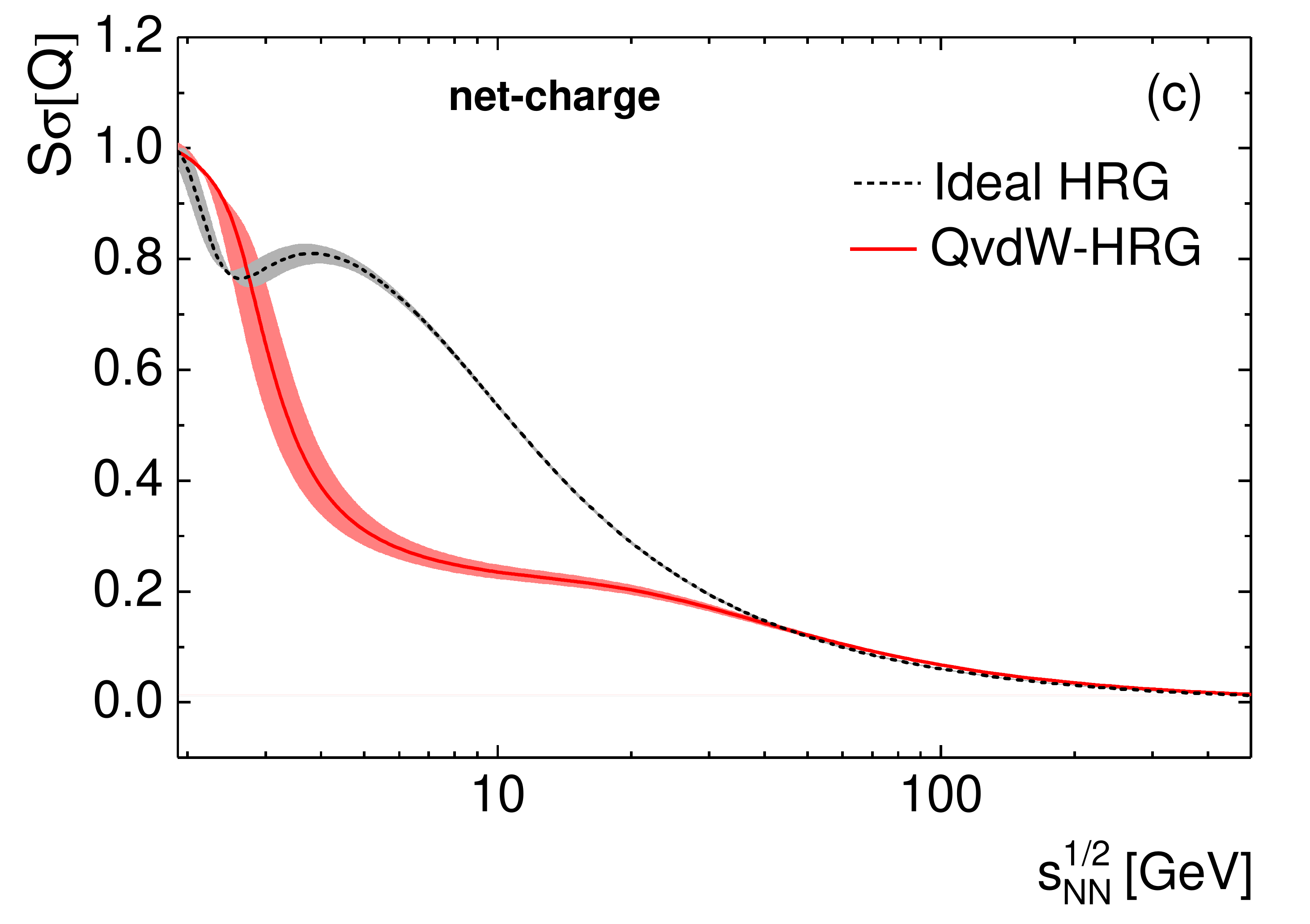}
\includegraphics[width=0.49\textwidth]{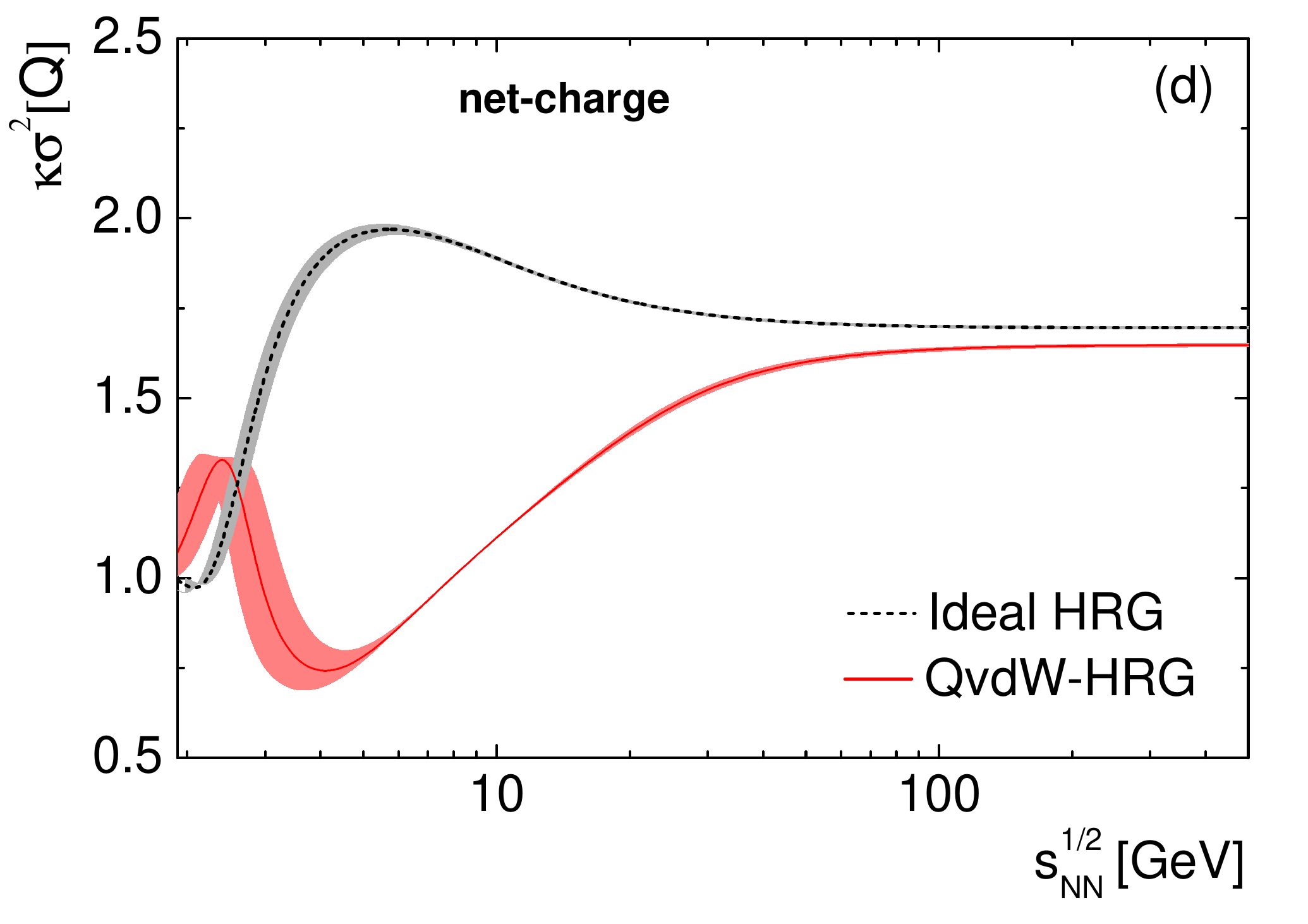}
\includegraphics[width=0.49\textwidth]{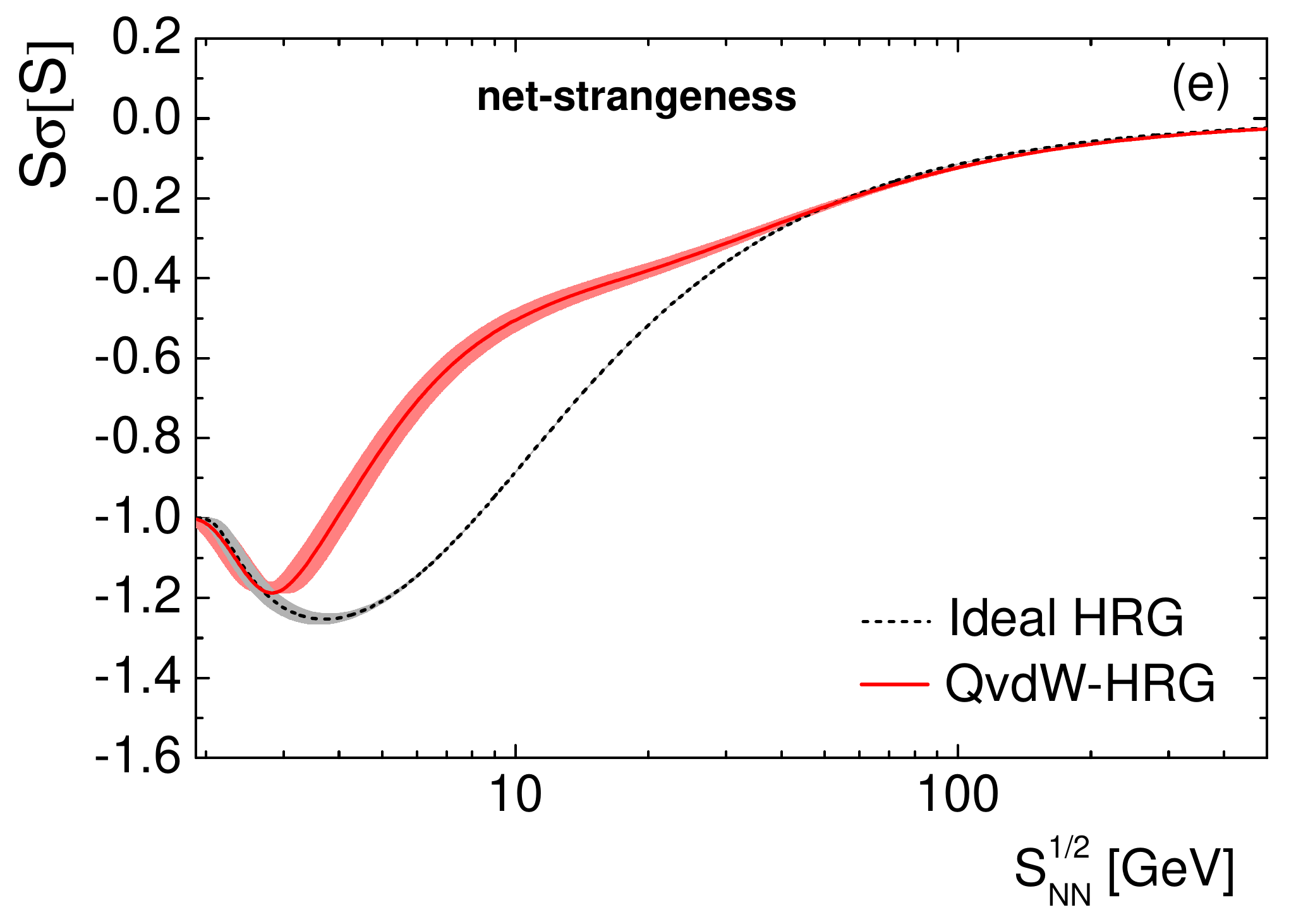}
\includegraphics[width=0.49\textwidth]{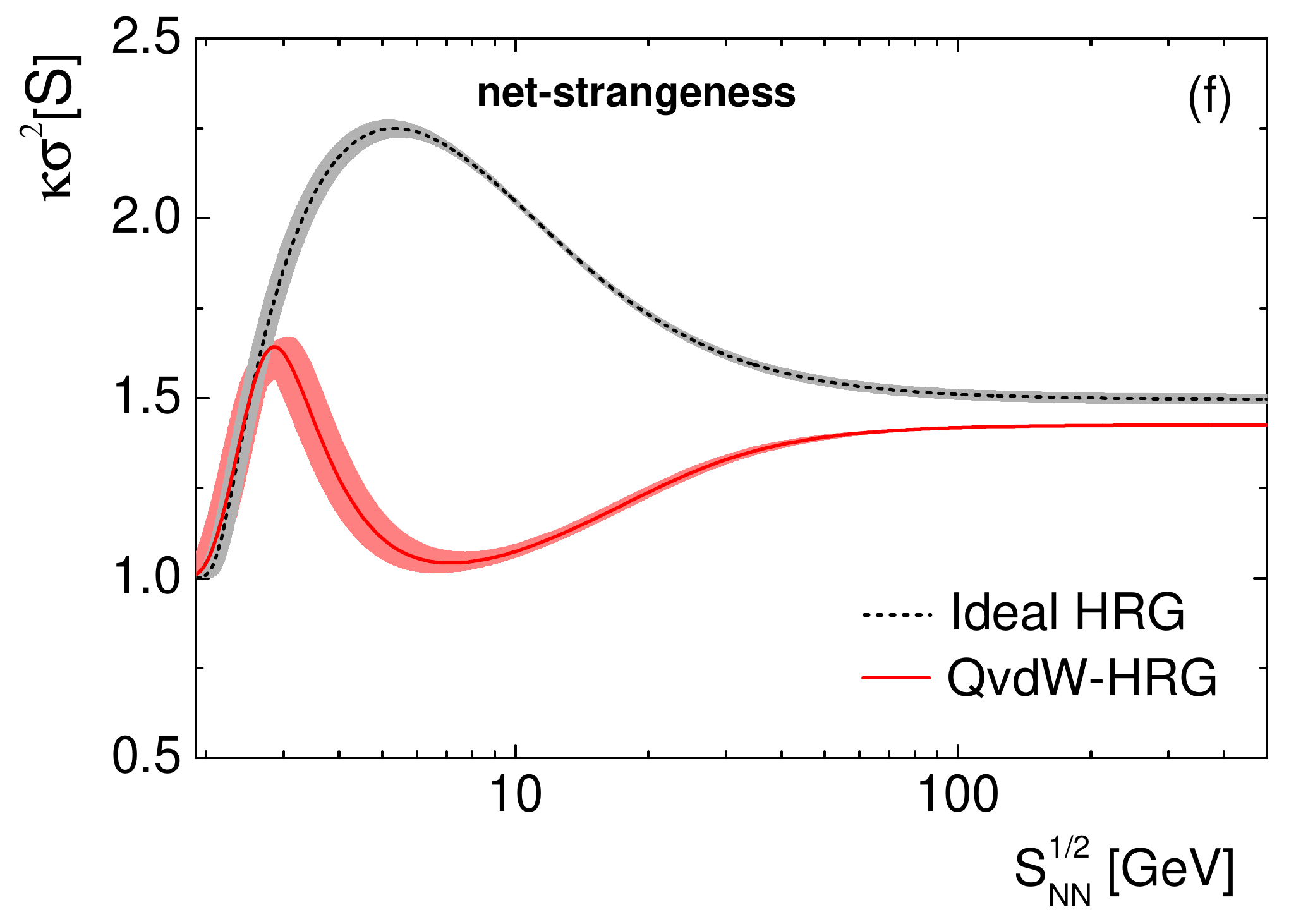}
\caption{\label{freeze-out} 
 Skewness ($lhs$) and kurtosis ($rhs$) of baryonic ($a,b$), electric ($c,d$), and strangeness ($e,f$) charge fluctuations 
within the IHRG- (dashed black curves)  and QvdW-HRG- (solid red curves) models, along the respective freeze-out lines of the models. 
The shaded areas along the curves represent the uncertainties of the predicted results due to the uncertainties of the fitted freeze-out line parameters.
Note that the predicted large differences of the fluctuation measures at BES~\rom{2}, GSI and FAIR energies, fully accessible experimentally to date. 
}
\end{figure}

The fluctuations exhibit a singular behavior at the CP: 
both the skewness and the kurtosis of charge fluctuations can approach the values $+\infty$, $-\infty$, or $0$, depending on the path with which the CP is approached in the phase diagram.
The strong influence of the CP on the higher moments of the distributions is apparent even far away from the CP. 
This is particularly true along the freeze-out lines.

 We have checked that the strangeness suppression effect due to $\gamma_s<1$ is rather small in the skewness and kurtosis of the net strangeness fluctuations. Therefore,
in calculations of charge fluctuations,
$\gamma_s$ is fixed to unity and the GCE is used. 
Figure \ref{freeze-out} shows the skewness
and the kurtosis of the charge fluctuations as functions of the center of mass energy $\sqrt{s_{NN}}$ as calculated in the IHRG and QvdW-HRG
models,
along the corresponding fitted chemical freeze-out lines~(\ref{Tmu}). 
The non-trivial behavior of higher order fluctuations measures superimposed on the shape of the freeze-out line leads to the complex, non-monotonic behaviour of these quantities as functions of the collision energy.
A pure asymmetric nuclear matter (the only constituents are nucleons) models both $S\sigma[B]$ and $\kappa\sigma^2[B]$ 
rather well
at moderate and low collision energies, $\sqrt{s_{NN}}\lesssim2.4~{\rm GeV}$. This energy corresponds to temperatures of $T\lesssim 60~{\rm MeV}$.
However, for quantitative description of the $S\sigma[Q]$ and the $\kappa\sigma^2[Q]$ values this approximation is not good enough, as, even at the lower energies, the contribution of the direct pions to the electric charge is quite substantial.
Accounting for both, nucleons and direct pions, does allow for a reasonable description of the skewness of the electric charge fluctuations, up to GSI energies, $\sqrt{s_{NN}}\lesssim2.4~{\rm GeV}$.
The deviations of the strangeness fluctuations at the intermediate beam energies from the IHRG baseline are mostly due to the contributions of strange baryons,  mainly $\Lambda$ and $\Sigma$,  which take part in the QvdW interactions.

 The results of the QvdW-HRG model for $\kappa\sigma^2[Q]$ and $\kappa\sigma^2[S]$ at high collision energies are close to the IHRG baseline.
This is due to the large contributions of pions and kaons, respectively, which are treated as non-interacting particles within QvdW-HRG approach.
In contrast, $\kappa\sigma^2[B]$ at high energies substantially deviates from the IHRG baseline due to interactions between (anti-)baryons.

Figures \ref{baryonic-fluct} and \ref{electric-fluct} demonstrate that the values of high-order charge fluctuations are highly sensitive to the location on the phase diagram. 
This sensitivity is strongest in proximity to the CP. 
In the QvdW-HRG model the fluctuations of the conserved charges in the system of  nucleons at small temperatures  are rather different in the dilute gaseous and dense liquid phases of interacting nucleons.
Within the QvdW-HRG model, the chemical freeze-out at the lowest $\sqrt{s_{NN}}$ takes place in the gaseous phase where the effects of the interactions are rather small.
Thus, strong deviations of the QvdW-HRG model fluctuations from the IHRG baseline appear mostly at intermediate beam energies where the baryonic density is the highest, see Fig.~\ref{fig-entropy1}~($a$).
Note that at present there is a lack of hadron multiplicity data in nucleus-nucleus collisions at low collision energies.
Our analysis show a strong  sensitivity of high-order charge fluctuations to the position of the chemical freeze-out line relative to the nuclear CP.  More data for hadron multiplicities in nucleus-nucleus reactions at low collision energies  are required to clarify the production of new hadrons as well as the final fractions of the light and intermediate  nuclear fragments.

 A quantitative comparison with the data
in heavy-ion collisions requires the appropriately chosen  acceptance region. This  region 
is defined  by the cuts in rapidity, transverse momentum, azimuthal angle, and other experimental limitations of measurements. For the event-by-event measurements in nucleus-nucleus reactions  the  $required$ $acceptance$ region  
should satisfy certain requirements. 
It should be a small 
part of the whole
phase space, thus,
the global charge conservation effects can be disregarded, and the statistical treatment within the grand canonical ensemble can be applied. On the other hand, this acceptance region
should be large enough to capture the relevant physics.

In order to reduce the volume fluctuation effects the most central collisions must be selected~\cite{Braun-Munzinger:2016yjz}.
Another way to reduce the volume fluctuation effects is to use the so-called strongly intensive quantities~\cite{Gorenstein:2011vq} as the fluctuation measures.
A detailed analysis of the required acceptance and volume fluctuation corrections is however outside the scope of the present paper and will be a subject of the future research.

 The baryon number fluctuations are calculated in the present paper with the assumption that all baryons are experimentally detectable.  This is not the case in reality. 
 Therefore, a binomial acceptance procedure, 
see Refs.~\cite{Kitazawa:2012at,Bzdak:2012ab,Vovchenko:2017ayq}, is usually  applied to account for this inability of the event-by-event measurements  of (anti)neutron numbers.
This procedure leads to an essential decrease of the observable baryon number fluctuations. In contrast to the baryon number, nearly all electric charges can be experimentally detected. Thus, our  results obtained for electric charge fluctuations are more suitable for a comparison with the experimental data.

Finally, an analysis of experimental data should be complemented with dynamical model simulations of heavy-ion collisions, where the effects of baryon-baryon interactions studied here are incorporated. 
The dynamical models can naturally incorporate the effects related to baryon number conservation and acceptance.
Some recent developments in this direction using transport models can be found in Refs.~\cite{He:2016uei,Ye:2018vbc}.

Note that the predicted large differences of the fluctuation measures at the high baryon density correspond to the center of mass energy regime from 3 to 30 GeV, 
which is readily accessible to the Beam Energy Scan \rom{2} run at Brookhaven National Laboratory, and at Helmholtzzentrum für Schwerionenforschung (GSI) HADES detector, as well as the Compressed Baryonic Matter (CBM) detector at  Facility for Antiproton and Ion Research (FAIR) and the Multi Purpose Detector (MPD) at NICA in Dubna.

\section{summary}
\label{sec-sum}
The quantum van der Waals hadron resonance gas model has been applied to study chemical freeze-out properties in heavy-ion collisions as well as the higher-order fluctuations of net baryon and net charge numbers.
The extracted chemical freeze-out parameters exhibit larger uncertainties as compared to the ideal hadron resonance gas model.
Similar to the IHRG model, the dependence of $T_{\rm ch}$ on $\mu_B^{\rm ch}$ in the QvdW-HRG model can be parametrized as a quartic polynomial in $\mu_B^{\rm ch}$, with parameters differing quite substantially from the IHRG case.
 Since both the LGPT and the chemical freeze-out are consistently obtained within a single model, their relative location is clarified.

 The beam energy dependences of the skewness and the kurtosis of the baryonic, electric, and strange charge fluctuations has been calculated along the obtained chemical freeze-out curve.
All six observables show large deviations from the ideal hadron resonance gas baseline at the highest baryon density at intermediate beam energies. 
These signals stem in the QvdW-HRG model from the nuclear critical point at $T \sim 20$~MeV and $\mu_B \sim 900$~MeV.
This observation must be taken into account in every experimental search for the QCD critical point in high energy nucleus-nucleus collision experiments using the higher order fluctuations of conserved charges.
This concerns in particular the Beam Energy Scan \rom{2} run at Brookhaven National Laboratory, as well as future GSI-HADES, FAIR-CBM, and NICA-MPD.

\section*{Acknowledgments}

The authors thank M. Gazdzicki, B. I. Lev,  and G. M. Zinovjev for fruitful discussions and useful comments. 
This research was supported by theme grant of department of physics and
astronomy of NAS of Ukraine: ``Dynamics of formation of spatially
non-uniform structures in many-body systems", PK 0118U003535.
H. St. appreciates the support through the Judah M.
Eisenberg Laureatus Chair at Goethe University, and the  Walter
Greiner Gesellschaft, Frankfurt.

\bibliography{main}

\end{document}